\documentclass[aps,preprint]{revtex4-1}
\usepackage{graphicx}
\usepackage{amsmath}
\usepackage{makeidx}
\usepackage{amsfonts}
\usepackage{amssymb}
\usepackage{dsfont}
\usepackage{color,soul}

\begin{document}

\title{Distance measures and evolution of polymer chains in their topological space}

\author{Alireza Mashaghi$^a$, Abolfazl Ramezanpour$^b$}
\affiliation{$^a$Harvard Medical School, Harvard University, Boston, Massachusetts, USA}
\affiliation{$^b$Department of Physics, University of Neyshabur, Neyshabur, Iran}
%\date{\today}

\begin{abstract}
Conformational transitions are ubiquitous in biomolecular systems, have significant functional roles and are subject to evolutionary pressures. Here we provide a first theoretical framework for topological transition, i.e. conformational transitions that are associated with changes in molecular topology. For folded linear biomolecules, arrangement of intramolecular contacts is identified as a key topological property, termed as circuit topology. Distance measures are proposed as reaction coordinates to represent progress along a pathway from initial topology to final topology. Certain topological classes are shown to be more accessible from a random topology. We study dynamic stability and pathway degeneracy associated with a topological reaction and found that off-pathways might seriously hamper evolution to desired topologies. Finally we present an algorithm for estimating the number of intermediate topologies visited during a topological reaction. The results of this study are relevant to, among others, structural studies of RNA and proteins, analysis of topologically associated domains in chromosomes, and molecular evolution.
\end{abstract}

%\pacs{} 
\maketitle

\section{Introduction}\label{S0}

Arrangement of the contacts in a folded linear polymer chain is a topological property of the chain. This arrangement, called circuit topology, has been rigorously defined for linear chains with intra-chain contacts and its use in studying the equivalence of folded molecular chains has been proposed \cite{MWT-structure-2014, MTM-pccp-2014}. Molecules can vary in size and sequence and yet have identical circuit topology. The topology framework allows for classification of chains into topological equivalence classes. This approach is applicable to biomolecular chains ranging from proteins, to RNA to complete chromosomes, all of which are folded linear chains held together by intra-chain contacts  \cite{MWT-structure-2014}. 

For two chains that do not fall within the same topological class, we lack a distance measure to quantify their differences. Distance measures are crucial for building molecular phylogeny based on topology and for describing topological dynamics of chains. An appropriate metric structure allows for the construction of an evolutionary landscape \cite{BC-pnas-1999}, probing of processes such as epistasis \cite{BKV-nature-2012}, and modeling formation and dynamics of topologically associated domains in chromosomes \cite{ZW-pnas-2015, JobDekker-nature-2012}. It can also serve as a reaction coordinate for conformational reactions such as folding \cite{CLW-pnas-2006}. Development of metric structures on a space of circuit topologies may have far reaching implications beyond polymer science. Similar linearly structured objects appear in areas other than biomolecular sciences \cite{FT-job-1963,DRBZ-natphys-2011,R-pre-2013}. These structures, though very different in nature, may share generic structural and dynamical properties.

For a simple model of a folded polymer chain, we define distance measures based on a set of reasonable local changes in the space of contact configurations, and use a minimum assignment algorithm to estimate one of the proposed distances. Because a topological treatment is agnostic to length, the model can be simplified by setting the length of every chain segment to any desired value, e.g. unity. The chain can then be readily represented by a graph in which the nodes correspond to the contact sites and the links represent intra-molecular interactions. The simple chain model allows us to disentangle topology from other structural features of molecules.

Despite the simplicity of our model, it is still a challenge to investigate topological dynamics of the chain and to identify conformational reaction pathways. Sampling from the exponentially large space of the possible pathways could be very time consuming for disordered and frustrated energy functionals of pathways, let alone the global constraint of connectivity in a pathway. An efficient way of sampling from such energy landscapes in sparsely (weakly) interacting systems is provided by the cavity method of statistical physics  \cite{MP-epjb-2001,MM-book-2009}, relying on the Bethe approximation. The recursive and local nature of these equations are exploited in approximate message-passing algorithms that have proven useful in the study of random constraint satisfaction and optimization problems \cite{MZ-pre-2002,MPZ-science-2002}. In particular, the cavity method could be very helpful when we have to deal with global or nonlocal constraints \cite{steiner-prl-2008,RRZ-epjb-2011,ABDZ-jstat-2013,YS-prl-2012,YSM-pnas-2013}, e.g. the connectivity constraint in the minimum-weight Steiner tree problem \cite{steiner-prl-2008}, and in the study of stochastic optimization problems \cite{ABRZ-prl-2011}, where computing the energy function is already hard. In all of these examples, it would be computationally expensive (if not impossible) to study large-scale problem instances with the standard optimization algorithms based on the Monte Carlo sampling.  

In this article, we map the entire space of contact configurations and study how two molecular configurations interconvert by the rearrangement of contacts. For this aim, we construct connectivity graphs of link configurations related by the above local changes used in the distance measures. Then, we employ an exhaustive search algorithm (for small systems) to study optimal evolution with respect to an appropriate energy functional of pathways in the configuration space. We also present an approximate message-passing algorithm for the optimal evolution problem in larger systems. To address the connectivity constraint of the evolution, we have to define some intermediate auxiliary variables, making the problem amenable to local message-passing algorithms. This allows us to find reasonable approximate solutions for the optimal paths connecting two boundary configurations.

\section{Definitions}\label{S1}
\begin{figure}
\includegraphics[width=16cm]{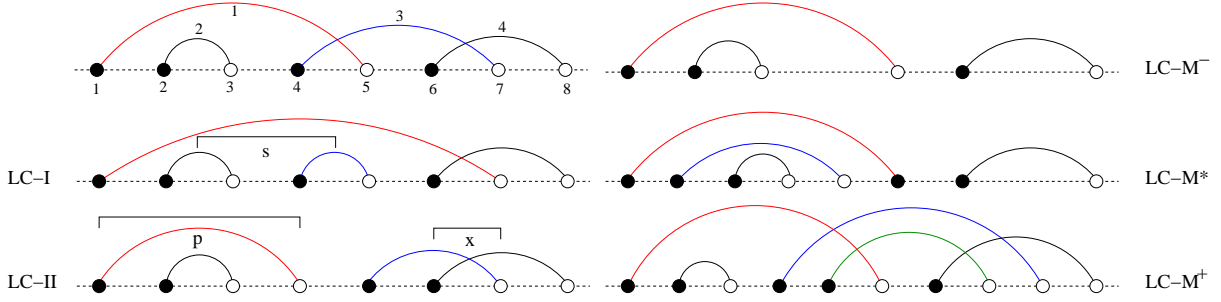} 
\caption{Illustrating the possible arrangements of contact pairs, and effects of different local changes on a link configuration. The links are ordered from left to right according to their first endpoint. Two links can be in parallel (p), series (s), and cross (x) states. The left panels show how a local change of type I (left-middle) and II (left-bottom) affect on links $1$ and $3$. The right panel shows:  (right-top) a local change of type $M^{-}$, where a contact is removed from the system, (right-bottom) a local change of type $M^{+}$, where a contact is created, and (right-middle) a local change of type $M^{*}=M^{+}M^{-}$, which is a combination of the above local changes.}\label{f1}
\end{figure}

Our graph representation of a chain with $M$ contacts includes $M$ links with endpoints $e_l=(i_l,j_l)$ labeled by $l=1,\dots,M$, where $i_l,j_l \in \{1,\dots,2M\}$ (Figure \ref{f1}). Here, a link configuration is defined by $\mathsf{L}=\{(i_l,j_l)|l=1,\dots,M\}$, where each endpoint takes part in one and only one contact. Note that $i_l \neq j_l$ and links are not directed $(i_l,j_l)\equiv (j_l,i_l)$. Any two links are in one of the three states: parallel ($p$), series ($s$), or cross ($x$) with respect to the backbone chain. The chain is directed from left to right $\{1,2,\dots,2M-1,2M\}$. We are interested in topologically different link configurations represented by an $M\times M$ matrix $\mathsf{A}_{l,l'} \in \{p,s,x\}$. For simplicity, here we do not care if such a configuration has a physical three dimensional realization.

A link configuration can also be represented by a perfect matching of the $2M$ endpoints $i=1,\dots,2M$. Here, the configuration is identified by a set of connectivity variables $\mathsf{C}\equiv \{c_{ij}=0,1|i<j\}$ showing the absence or presence of connections between the endpoints.
Each perfect matching defines a class of topologically equivalent link configurations related by a permutation of the link labels. The number of such perfect matchings is $(2M-1)!! \equiv (2M-1)\times(2M-3)\dots \times 1$, and for each one there are $M!$ ways of labeling the links. In other words, there are $M!$ matrices $\mathsf{A}$ representing the set of topologically equivalent link configurations. 
Given a link configuration, one can easily construct the unique matrix $\mathsf{A}$; the endpoints $e_l=(i_l,j_l)$ and $e_{l'}=(i_{l'},j_{l'})$ are enough to identify the element $\mathsf{A}_{l,l'}(e_l,e_{l'}) \in \{p,s,x\}$.

A link configuration of $M$ links has $N=M(M-1)/2$ pairs of links that can be partitioned into three disjoint subsets of size $N_p,N_s,N_x$ depending on their relative state $p,s,x$. The links can have an arbitrary labeling, but a convenient one that we will use later is the one in which the links are labeled from left to right according to the order of their first endpoints. At some point we will consider structured or modular link configurations with groups of links organized in well separated communities or components. We will consider simple modules of $m$ links with all the link pairs of type $q=p,s,x$, represented by $qm$.
For example, $q_1m_1q_2m_2$ shows a configuration with two modules of types $q_1,q_2$ and number of links $m_1,m_2$, respectively. 
Figure \ref{f1} displays the main definitions and notations we use in this paper.

\section{Distance measures}\label{S2}
Given a link configuration, one can change the link arrangement in different ways to obtain a nearby configuration.
In this study, we will use the following local changes to construct a connectivity graph $\mathcal{G}$ of configurations and to define appropriate distance measures in the space of contact configurations.

(I) Consider two links $l$ and $l'$ with endpoints $e_l=(i_l,j_l)$ and $e_{l'}=(i_{l'},j_{l'})$, respectively. From this we construct the other two topologically distinct configurations: $e_l=(i_l,i_{l'}), e_{l'}=(j_l,j_{l'})$ and $e_l=(i_l,j_{l'}), e_{l'}=(j_l,i_{l'})$, obtained by an interchange of the endpoints. We call this a local change of type I (LC-I). This changes not only $\mathsf{A}_{l,l'}$ but may also change other matrix elements $\mathsf{A}_{l,l''}$ and $\mathsf{A}_{l'',l'}$. As a result, the numbers $N_{p,s,x}$ may considerably change by a LC-I.     

(II) Consider two neighboring endpoints $(k,k+1)$ on the chain belonging to two different links, say $e_l=(i_l,j_l=k)$ and $e_{l'}=(i_{l'}=k+1,j_{l'})$. Then we change the order of the neighboring endpoints to obtain $(i_l,j_l=k+1)$ and $(i_{l'}=k,j_{l'})$, see Fig. \ref{f1}.  We call this a local change of type II (LC-II). This only changes the matrix element $\mathsf{A}_{l,l'}$. Here, the numbers $N_{p,s,x}$ change smoothly (at most by one).

\begin{figure}
\includegraphics[width=16cm]{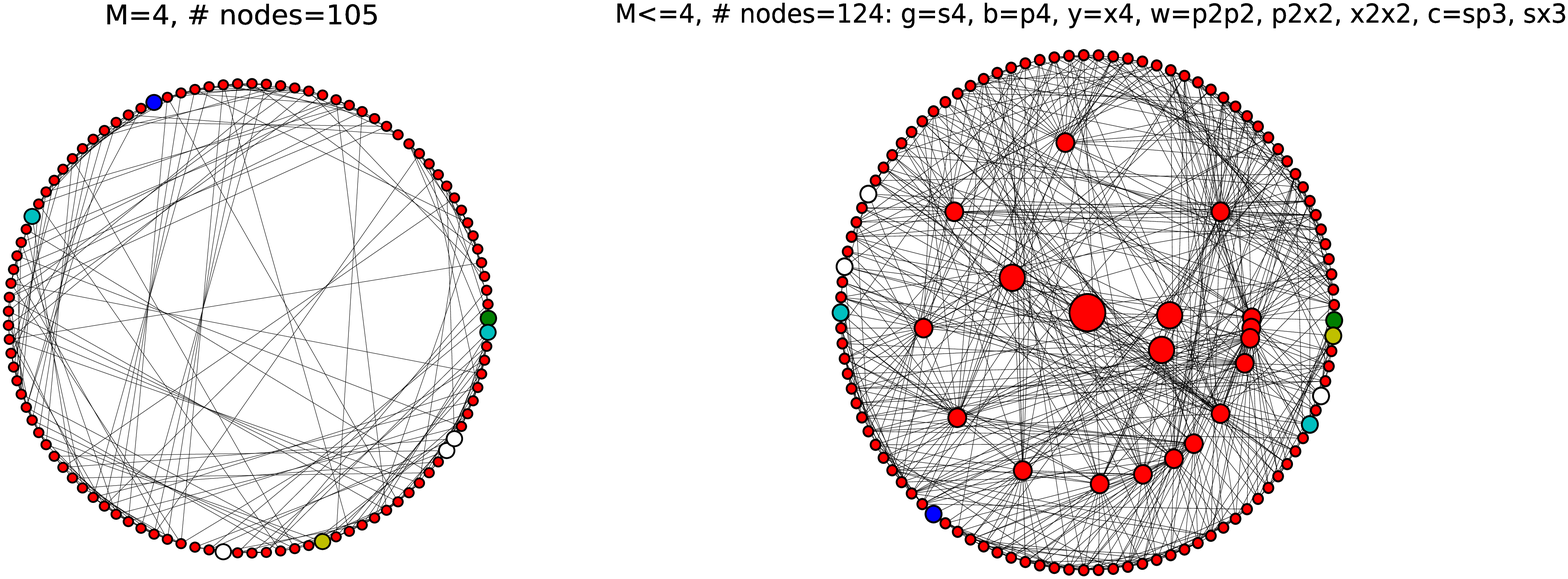} 
\caption{The graph representation of link configurations (shown by nodes) for $M=4$ (left) and $M\le 4$ (right). Two nodes in the left panel are connected if the corresponding link configurations are related by a local change of type II.  A connection in the right panel means the two configurations are related by LC-II or LC-$M^{\pm}$. The special link configurations are:  $s4$ (green), $p4$ (blue), $x4$ (yellow), $p2p2, p2x2,x2x2$ (white), and $sp3$, $sx3$ (cyan). The bigger nodes inside the circle in the right panel show configurations with a smaller number of links.}\label{f2a}
\end{figure}

Starting from a link configuration we can obtain all the other ones from permutations of the endpoints $i=1,2,\dots,2M$ and the link labels $l=1,2,\dots,M$; there are $2^MM!$ permutations of the endpoints and the link labels that lead to a topologically equivalent configuration. In fact, starting from an arrangement of the endpoints we can reach to any other one by a sequence of elementary transpositions swapping two adjacent endpoints. Thus, the space of link configurations is connected under the LC-II updates. Figure \ref{f2a} shows a small connectivity graph $\mathcal{G}_{II}$ of link configurations related by the LC-II. We see that there is no odd loop in this graph; we always need an even number of local changes of type II to return to a given link configuration.

Note that the number of links $M$ is fixed in the above local changes. But we may consider the case in which one link can be removed from or added to the system. We say two link configurations (with different number of links) are connected by a local change of type $M^{\pm}$ if the larger configuration (with larger number of links) can be obtained by adding one link to the other configuration.      
The process in which one link is removed from the system (LC-$M^-$), and is replaced with another link connecting two new endpoints (LC-$M^+$), defines another local change (LC-$M^{*}$) in the subspace of fixed $M$. Note that each endpoint belongs to one and only one link. When we add (remove) a contact we also add (remove) the two associated endpoints. Moreover, we could also allow for different local changes to happen; for example, two link configurations can be related by either LC-II or LC-$M^{\pm}$. Figure \ref{f2a} shows such an extended connectivity graph of link configurations $\mathcal{G}_{II+M^{\pm}}$ with $M\le 4$ links.

Based on the above local changes we define the following distance measures:

(DI) Given the connectivity patterns $\mathsf{C}_1,\mathsf{C}_2$ of the endpoints in two link configurations, we can easily compute the Hamming distance $D_{I}(\mathsf{C}_1,\mathsf{C}_2)=\sum_{i<j} (1-\delta_{c_{ij}^1, c_{ij}^2})/4$ of the two configurations. Here $\delta_{c,c'}=1$ if $c=c'$, otherwise $\delta_{c,c'}=0$. A local change of type I increases this distance by one, but it could result to considerable changes in other macroscopic properties of the chain. For instance, the number of links in the shortest path connecting the two ends of the chain (end-to-end distance) is very sensitive to a LC-I. In other words, there are very close link configurations (according to this measure) with very different end-to-end distances. The situation is of course better for the average shortest distance (the minimum number of links needed to go from one endpoint to another one). 

\begin{figure}
\includegraphics[width=16cm]{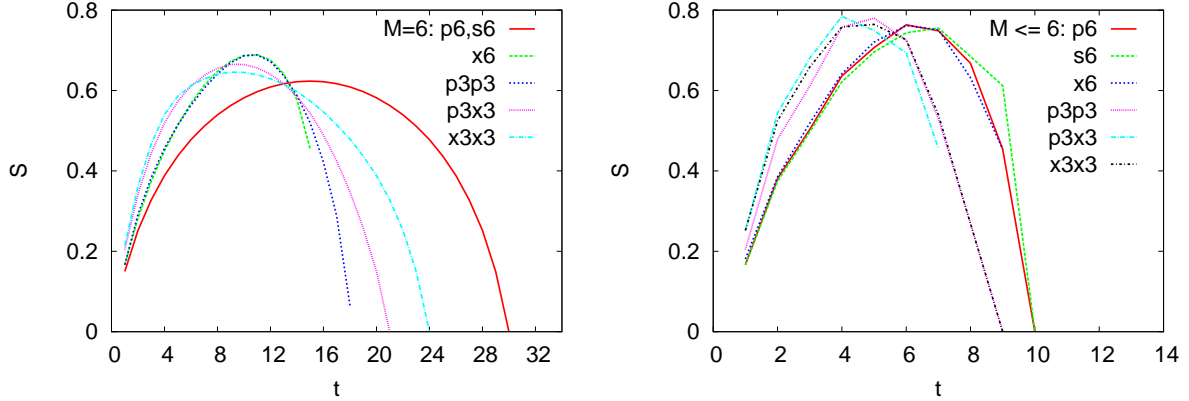} 
\caption{The exact entropy $S(t)$ (logarithm of the number of link configurations) at distance $t$ from some reference configurations ($p6,s6,x6,p3p3,p3x3,x3x3$). In the left panel, $t$ shows the number of local changes of type II connecting two link configurations in the subspace of fixed $M=6$. In the right panel, $t$ shows the number of local changes of type II and $M^{\pm}$ connecting two link configurations with possibly different number of links $M\le 6$.}\label{f2b}
\end{figure}

(DII) The local changes of type II suggest another distance measure $D_{II}(\mathsf{L}_1,\mathsf{L}_2)$ as the minimum number of LC-IIs we need to go from configuration $\mathsf{L}_1$ to $\mathsf{L}_2$. Both the end-to-end and the average shortest path distances are more correlated with this measure than the previous one. Notice that there is no odd loop in $\mathcal{G}_{II}$; any two link configurations connected by a path of even length can not be connected by another path of odd length. Therefore, all the paths connecting the same boundary configurations have even (odd) lengths. 
The number of link configurations at distance $t$ from a reference configuration $\mathcal{N}(t)$ shows how the other configurations are distributed around the configuration. Figure \ref{f2b} shows the entropy $S(t)\equiv (\ln \mathcal{N}(t))/(M\ln M)$ obtained by an exhaustive enumeration algorithm for $M=6$ links. It is observed that structured configurations like $x3x3$ (with two $x3$ modules) are closer to the other configurations than the all-$s$ ($s6$) and the all-$p$ ($p6$) configurations. As the figure shows, the difference becomes clearer in the extended connectivity graph, where two link configurations are directly connected if they are related by LC-II or LC-$M^{\pm}$.

\subsection{Distance of two matrix configurations}\label{S21}
The Hamming distance of two link configurations represented by matrices $(\mathsf{A}^1,\mathsf{A}^2)$ is simply  
$D(\mathsf{A}^1,\mathsf{A}^2)=\sum_{l<l'} (1-\delta_{\mathsf{A}_{ll'}^1, \mathsf{A}_{kk'}^2})$ for a given assignment $k(l)$ of the link labels in the two matrices. Here, to ease the notation, we used $k,k'$ for $k(l),k(l')$. When the links are indistinguishable and the contact labels are irrelevant, we define the distance measure
\begin{align}
D_{III}(\mathsf{A}^1,\mathsf{A}^2)=\min_{assignment}\left\{ \sum_{l<l'} (1-\delta_{\mathsf{A}_{ll'}^1, \mathsf{A}_{kk'}^2}) \right\}.  
\end{align}
If two matrix configurations $\mathsf{A}^1,\mathsf{A}^2$ are related by a LC-II then $D_{III}(\mathsf{A}^1,\mathsf{A}^2)=1$. We stress that in contrast to the other two measures, here the link and contact labels are not important; $D_{III}(\mathsf{A}^1,\mathsf{A}^2)=0$ for any two link configurations that have the same number of modules with similar internal structures irrespective of their ordering.

Finding the above minimum distance is a minimum assignment problem: we are to find a one-to-one assignment $\mathbf{l} \to \mathbf{k}$ of the link labels in the two configurations minimizing the Hamming distance $D(\mathsf{A}^1,\mathsf{A}^2)$. Here, we briefly describe an approximate algorithm to find such an assignment. More details are given in Appendix \ref{MD-app}.

We consider the probability measure $\mu(\mathbf{k}) \propto e^{-\beta D(\mathsf{A}^1,\mathsf{A}^2)}$ of assignment $\mathbf{k}$, where $D(\mathsf{A}^1,\mathsf{A}^2)$ plays the role of an energy function, and $\beta$ is an inverse temperature controlling the typical Hamming distances. Then, we use Bethe approximation to compute the local probability marginals $\mu_l(k)$ of assigning $l \to k$. In the Bethe approximation, the local marginals are written in terms of the cavity probability marginals $\mu_{l'\to l}(k')$ of assigning $l'\to k'$ in the absence of link $l$. For a moment, suppose the interaction (dependency) graph of the variables, defined by the energy function, is a tree. Then, the cavity marginal $\mu_{l'\to l}(k')$ can be written in terms of the other cavity marginals received from the neighboring variable nodes except $l$. The recursive equations governing these cavity marginals are called the belief propagation (BP) equations \cite{MM-book-2009,KFL-inform-2001}. For arbitrary interaction graphs, the BP equations can still be used to find good estimations for the local probability marginals. The quality of this approximation then depends on the structure and strength of the interactions. In our model, the approximated BP equations for the cavity marginals are given by       
\begin{align}
\mu_{l'\to l}(k') \propto \prod_{l''\neq l,l'}\left( \sum_{k'' \neq k'} e^{-\beta (1-\delta_{\mathsf{A}_{ll'}^1, \mathsf{A}_{kk'}^2})} \mu_{l''\to l'}(k'')\right).
\end{align}
We obtain the local marginals $\mu_l(k)$ in the same way, but considering all the incoming cavity marginals $\mu_{l'\to l}(k')$.  The limit $\beta \to \infty$ of the local marginals would be enough to find an approximate solution by a decimation algorithm, as explained in Appendix \ref{MD-app}. 

We used the above algorithm to find an approximate assignment minimizing the Hamming distance of two randomly generated link configurations. For $M=10,20,30,40$ links we find the following estimations of the minimum distances $2D_{III}/(M(M-1))=0.371\pm 0.008, 0.316\pm 0.006, 0.294\pm 0.007,0.289\pm 0.006$, respectively.

\section{Minimum evolution problem}\label{S3}
Consider an evolution path of length $T$ starting from link configuration $\mathsf{L}_0$, ending up at configuration $\mathsf{L}_T$, and connecting neighboring link configurations. Define $\mathcal{A}_{\mathsf{L},\mathsf{L}'}=1$ for two neighboring link configurations related by a local change, otherwise $\mathcal{A}_{\mathsf{L},\mathsf{L}'}=0$. The aim is to find an optimal evolution minimizing an energy functional of the path, $\mathcal{E}[\mathsf{L}_1,\dots, \mathsf{L}_{T-1}]$, depending on the intermediate link configurations. To this end, we need to sample the space of possible pathways with the following dynamical partition function: 
\begin{align}
\mathcal{Z}(\mathsf{L}_0\to \mathsf{L}_T)=\sum_{\mathsf{L}_1,\mathsf{L}_2,\dots, \mathsf{L}_{T-1}} e^{-\beta \mathcal{E}[\mathsf{L}_1,\dots, \mathsf{L}_{T-1}]}\mathcal{A}_{\mathsf{L}_T,\mathsf{L}_{T-1}} \dots \mathcal{A}_{\mathsf{L}_{t+1},\mathsf{L}_t} \dots \mathcal{A}_{\mathsf{L}_1,\mathsf{L}_0}.
\end{align}
Each path has the statistical weight $e^{-\beta \mathcal{E}[\mathsf{L}_1,\dots, \mathsf{L}_{T-1}]}$ depending on the path energy and the inverse temperature parameter $\beta$. At the end, we will need to take the limit $\beta\to \infty$ to focus on the optimal pathways. 

We assume that the energy functional can be written as $\mathcal{E}=\sum_{t=1}^{T-1}E(t)+ \sum_{t=1}^{T}E(t-1,t)$, where $E(t)$ depends on the link configuration at time step $t$, and $E(t-1,t)$ is a function of the transformation from step $t-1$ to $t$. More specifically, we will take $E(t)=-N_p(t)$, and $E(t-1,t)=-\sum_{q<q'}\lambda_{q\to q'}N_{q\to q'}(t-1,t)$ for $q=p,s,x$. Here $N_{q\to q'}$ is the number of contact pairs changed from $q$ to $q'$. 
Simple folding models show that contact configurations with larger $N_p$ exhibit smaller folding times \cite{MTM-pccp-2014}. That is why we choose an energy functional that decreases with the number of parallel contact pairs.

The following study can also be done with more general energy functions, for instance, $E(t)=\sum_r \lambda(r) M(r;t)+\sum_{q=p,s,x}\sum_{d} \lambda_q(d)N_q(d;t)$. Here $M(r;t)$ is the number of links of length $r\equiv |j_l-i_l|$ at time step $t$, and $N_q(d;t)$ is the number of contact pairs of type $q=p,s,x$ having distance $d\equiv |i_l-i_{l'}|$. We observed that such energy functions could be  useful in the study of complex link configurations in the presence of structural modules and sectors \cite{MR-RSC-2015}.

\begin{figure}
\includegraphics[width=16cm]{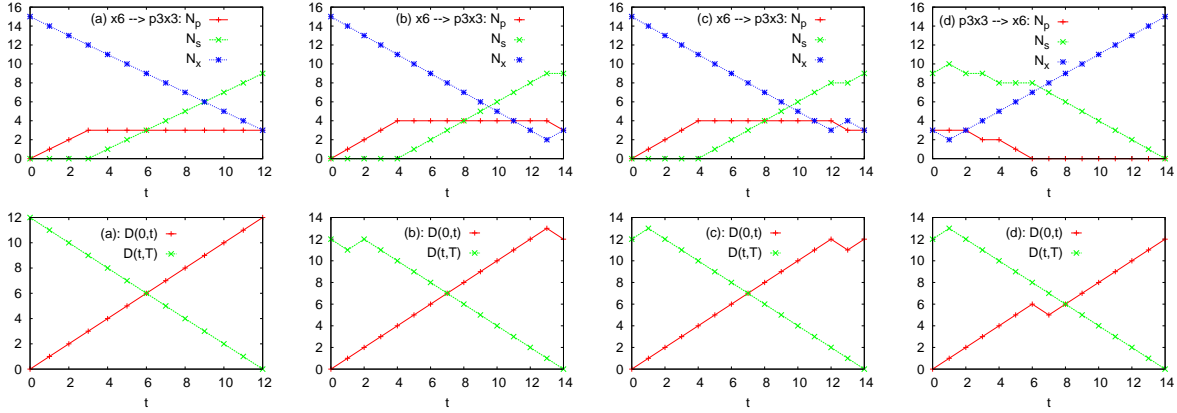} 
\caption{Evolution with local changes of type II: $N_{p,s,x}(t)$, and $D_{II}(t)$ of the intermediate link configurations from the boundary configurations in the paths obtained by the exact algorithm for $M=6$ links from the all-$x$ ($x6$) configuration at $t=0$ to a modular structure of two components ($p3x3$) at shortest distance $D_{II}=12$. Besides the shortest path (a), we display the optimal paths for $T=14$ maximizing $\mathcal{N}_p$ (b), and the path maximizing $\mathcal{N}_{x\to p,s}$ for $x6\to p3x3$ (c) and $p3x3\to x6$ (d). Here $t$ denotes the number of local changes of type II. The path degeneracy $g$ and energy gap $\Delta$ are: $(g=1008,\Delta=31)_a$, $(g=2688,\Delta=1)_b$, $(g=32904998,\Delta=14)_c$, $(g=32409638,\Delta=2)_d$.}\label{f3}
\end{figure}

For now, we consider simple paths with no loops, that is each configuration in the path is visited only once.
Figure \ref{f3} displays the optimal evolutions (obtained by an exhaustive search algorithm) connecting the $x6$ and $p3x3$ configurations by the local changes of type II, for a small number of links ($M=6$). Besides, we report the degeneracy $g$ (number of optimal paths) and the energy gap $\Delta$ between the optimal and the next optimal paths. These quantities provide a measure for stability of the optimal path. As the figure shows, for $M=6$ links and $T=12$, the optimal path from $x6 \to p3x3$ maximizing $\mathcal{N}_p\equiv \sum_{t=1}^{T-1}N_p(t)$ has degeneracy $g=1008$ and energy gap $\Delta=31$. Increasing the evolution time to $T=14$, results in $(g=2688,\Delta=1)$. In addition, we observe that the optimal paths maximizing the number of transitions $\mathcal{N}_{x\to p,s} \equiv \sum_{t=1}^{T}[N_{x\to p}(t-1,t)+N_{x\to s}(t-1,t)]$ have a very large degeneracy compared to those that maximize the total number of parallel contact pairs.
In Appendix \ref{figs-app}, we give other examples of evolution with the other local changes, connecting also link configurations with different number of links.

\begin{figure}
\includegraphics[width=14cm]{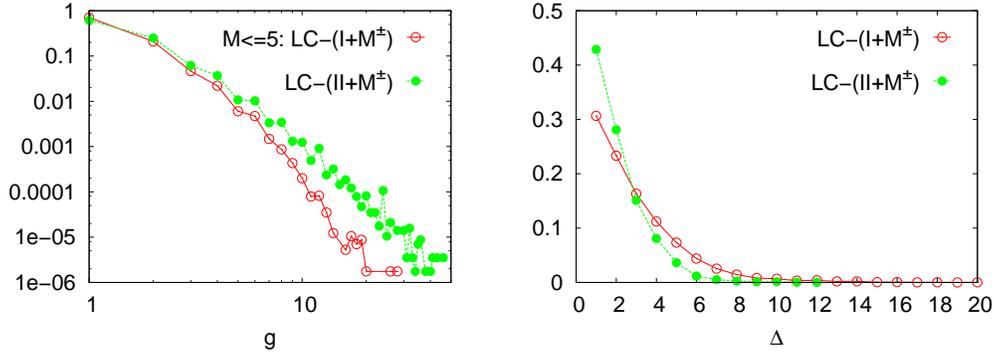} 
\caption{Probability distribution of the degeneracy $g$, and gap between the optimal and the next optimal paths $\Delta$, for shortest paths maximizing $\mathcal{N}_p$ in the connectivity graph of link configurations with $M\le 5$ links using the LC-(I+$M^{\pm}$) and LC-(II+$M^{\pm}$).}\label{f4}
\end{figure}

Next, we study some statistical properties of the optimal paths in the connectivity graph of link configurations with $M\le 5$ links. To this end, we take the adjacency graphs obtained by different choices of the local changes, and compare the number of optimal shortest paths maximizing $\mathcal{N}_p$, with a given path length $T$, degeneracy $g$, and energy gap $\Delta$. Table \ref{table-0} gives the average value $\langle o\rangle$ and standard deviation $\sigma_o=\sqrt{\langle o^2\rangle-\langle o\rangle^2}$ of these quantities, in addition to values for the correlation coefficients $r(o,o')=(\langle oo' \rangle-\langle o \rangle\langle o' \rangle)/(\sigma_{o}\sigma_{o'})$. Longer optimal paths are expected to have larger degeneracies and smaller gaps; that would result in positive and negative values for the correlation coefficients $r(T,g)$ and $r(T,\Delta)$, respectively. As the table shows, this does note always happen, for example, for local changes of type II+$M^*$. We also observe that the local changes of type II behave differently from the other local changes, with very strong fluctuations in $g$ and $\Delta$, along with very small correlation coefficients. In Fig. \ref{f4}, we also display the probability distribution of $g$ and $\Delta$ for the optimal shortest paths in the connectivity graphs $\mathcal{G}_{I+M^{\pm}}$ and $\mathcal{G}_{II+M^{\pm}}$. Here, one may prefer LC-(I+$M^{\pm}$) to LC-(II+$M^{\pm}$), as the optimal paths have in average smaller degeneracy and larger energy gap in the former than the latter case.

\begin{table}
\begin{center}

\begin{tabular}{|c||c|c|c|c|c|c|}
  \hline

& $(\langle T\rangle,\sigma_T)$ &  $(\langle g\rangle,\sigma_g)$ & $(\langle \Delta \rangle,\sigma_{\Delta})$ &  $r(T,g)$ & $r(T,\Delta)$& $r(g,\Delta)$ \\

  \hline

    LC-I     & $(3.216,0.770)$  & $(1.835,1.483)$  & $(4.192,4.343)$ & $0.264$ & $-0.09$ & $0.102$ \\

  \hline

    LC-II    & $(6.134,2.129)$  & $(4.939,61.762)$  & $(7.937,10.608)$ & $0.037$ & $0.022$  & $0.041$ \\

  \hline

    LC-(I+$M^{\pm}$)    & $(2.760,0.623)$  & $(1.439,0.889)$  & $(2.937,2.209)$ & $0.227$ & $-0.068$ & $-0.096$\\

  \hline

    LC-(II+$M^{\pm}$)    & $(3.418,0.924)$  & $(1.684,1.371)$  & $(2.102,1.339)$ & $0.291$ & $-0.175$ & $-0.08$\\

  \hline

    LC-(I+$M^*$)    & $(2.374,0.572)$  & $(1.451,0.940)$  & $(3.636,2.844)$ & $0.225$  & $0.021$ & $-0.005$ \\

  \hline

    LC-(II+$M^*$)      & $(2.697,0.667)$  & $(1.579,1.134)$  & $(4.416,3.694)$ & $0.217$ & $0.121$ &  $0.021$ \\

  \hline

\end{tabular}

\vskip 0.5cm

\caption{Statistical properties of the optimal shortest paths maximizing $\mathcal{N}_p$ in the connectivity graphs of link configurations with $M\le 5$ links obtained by different local changes: I, II, I+$M^{\pm}$, II+$M^{\pm}$, I+$M^*$, II+$M^*$. Here $T, g$, and $ \Delta$ denote the path length, the degeneracy of the optimal path, and the energy gap between the optimal and the next optimal paths, respectively. The data in case LC-II are restricted to paths of length $T\le 12$ to reduce the computation time. The average and standard deviation of variable $o$ are denoted by $\langle o \rangle$ and $\sigma_o$. The correlation coefficient of two variables $o,o'$ is computed by $r(o,o')=(\langle oo' \rangle-\langle o \rangle\langle o' \rangle)/(\sigma_{o}\sigma_{o'})$.}\label{table-0}
\end{center}
\end{table}

Consider two link configurations $\mathsf{L}_0,\mathsf{L}_T$ connected by a sequence of local changes, that is $\mathsf{L}_T=\mathbf{u}_T\cdots \mathbf{u}_1\mathsf{L}_0$. There could be different orderings of the LCs connecting the same boundary configurations. The question is how these different orderings affect a macroscopic behavior (phenotype) of the chain, for example a monotonically increasing (fitness) function of the $N_{p,s,x}$. For simplicity, here we focus on the case of two local changes $\mathbf{u},\mathbf{v}$ of type II. We say the local change $\mathbf{u}$ commutes with $\mathbf{v}$ in the context of $\mathsf{L}$ if $\mathbf{v}\mathbf{u}\mathsf{L}=\mathbf{u}\mathbf{v}\mathsf{L}$. In table I of Appendix \ref{ordering-app}, we summarize the possible effects of two commutative local changes on a contact configuration. As the table shows, the changes in numbers $N_{p,s,x}$ in the two paths connecting $\mathsf{L}$ to $\mathsf{L}'=\mathbf{v}\mathbf{u}\mathsf{L}=\mathbf{u}\mathbf{v}\mathsf{L}$ are correlated depending on how that quantity changes from $\mathsf{L}$ to $\mathsf{L}'$. In particular, when $N_{q}$ increases (or decreases) the corresponding changes in the two paths can not have different signs. 

The optimal paths we obtained so far were simple with no loops (also called off-pathways). The off-pathways (if allowed) can localize the dynamics in a small region of the configuration space wasting the evolution time. To escape from these traps, one may increase the path length $T$ or somehow disturb the system in the hope of finding another path dominating the off-pathways, but probably another set of off-pathways would appear. See Table II in Appendix \ref{figs-app}, for some examples of evolution in the presence of off-pathways.

\subsection{An approximate evolution algorithm}\label{S31}
For larger number of links, one has to think of other approximate algorithms. To be specific, in the following we consider only evolutions with the local changes of type II. We will need to introduce some auxiliary variables to represent the global constraint of path connectivity in terms of local constraints amenable to local message-passing algorithms. Then, we will apply Bethe approximation to obtain an efficient way of dealing with the above minimum evolution problem.     

Let us label the links from left to right according to the order of their first endpoints. To shorten the evolution time, we allow for more than one LC-II in each step of the evolution and represent the transformation from  $\mathsf{L}_{t-1}$ to $\mathsf{L}_{t}$ by a matching of the links $\mathbf{u}(t)$; links $l$ and $l'$ are involved in a local change of type II if $u_{ll'}(t)=1$, otherwise $u_{ll'}(t)=0$. Moreover, $u_{ll'}(t)$ could be $1$ only if links $l$ and $l'$ have neighboring endpoints on the chain. The matching property of $\mathbf{u}(t)$ means that if $u_{ll'}(t)=1$ then $u_{ll''}(t)=u_{l'l''}(t)=0$ for all $l'' \neq l,l'$. This property ensures that order of the local changes in one step is not important, therefore, we can uniquely determine $e_l(t),e_{l'}(t)$ from $e_k(t-1),e_{k'}(t-1)$ in case $u_{ll'}(t)=1$. If necessary, we also interchange the labels $(k,k') \to (l=k',l'=k)$ to ensure that in each step of the evolution the links are ordered according to their first endpoints.

The statistical properties of the problem can be obtained from the following dynamical partition function
\begin{align}
\mathcal{Z}(\mathsf{L}_0\to \mathsf{L}_T)=\sum_{\mathbf{u}(1),\mathbf{u}(2),\dots, \mathbf{u}(T)} e^{-\beta \mathcal{E}} \delta_{\mathsf{L}(\mathbf{u}(1),\mathbf{u}(2),\dots,\mathbf{u}(T)|\mathsf{L}_0),\mathsf{L}_T}.
\end{align}
Note that given the initial configuration $\mathsf{L}_0$ and transformations $\{\mathbf{u}(1),\mathbf{u}(2),\dots,\mathbf{u}(T)\}$, we can uniquely construct the configuration at step $t$, that is $\mathsf{L}_t=\mathsf{L}(\mathbf{u}(1),\mathbf{u}(2),\dots,\mathbf{u}(t)|\mathsf{L}_0)$. Here $\delta_{\mathsf{L},\mathsf{L}_T}$ is one if $\mathsf{L}=\mathsf{L}_T$, otherwise it is zero.
   
The above problem can be solved by a dynamic programming algorithm working with the cavity marginals $\mu_{t\to t+1}(\mathsf{L}_t)$ and $\mu_{t\to t-1}(\mathsf{L}_t)$. These are the probability of having configuration $\mathsf{L}_t$ in the absence of the energy terms and constraints in the other segment of the pathway; i.e. $(t,T]$ for the forward message $\mu_{t\to t+1}$, and $[0,t)$ for the backward message $\mu_{t\to t-1}$. More precisely, 
\begin{align}\label{ebp}
\mu_{t\to t+1}(\mathsf{L}_t) \propto e^{-\beta E(t)}\sum_{\mathbf{u}(t)} \delta_{\mathsf{L}(\mathbf{u}(t)|\mathsf{L}_{t-1}),\mathsf{L}_t} e^{-\beta E(t-1,t)}\mu_{t-1\to t}(\mathsf{L}_{t-1}),
\end{align}
and similarly for $\mu_{t\to t-1}(\mathsf{L}_t)$.

Note that working with an exact representation for the cavity marginals is computationally very expensive for large problem sizes. Thus, we have to resort to reasonable approximations working with an efficient and succinct representation of the cavity messages; see \cite{LSS-prb-2008,AM-jstat-2011,LMOZ-pre-2014} for examples in the quantum and dynamic cavity methods. Here we explain the main approximations used in this study. The reader can find more details in Appendix \ref{ME-app}. 
         
We represent a link configuration $\mathsf{L}$ by the set of endpoints $e_l=(i_l,j_l)$ and approximate the cavity marginals by a Bethe distribution \cite{LSS-prb-2008}, for instance,
\begin{align}
\mu_{t-1\to t}(\mathsf{L}) \approx \prod_l \mu_{t-1\to t}^l(e_l) \prod_{l<l'} \frac{\mu_{t-1\to t}^{ll'}(e_l,e_{l'})}{\mu_{t-1\to t}^l(e_l)\mu_{t-1\to t}^{l'}(e_{l'})}.
\end{align}
Here $\mu_{t-1\to t}^{l}(e_l)$ and $\mu_{t-1\to t}^{ll'}(e_l,e_{l'})$ are the one-link and two-link marginals of $\mu_{t-1\to t}(\mathsf{L})$.  
Using this form of the cavity messages in the right hand side of Eq. \ref{ebp}, we employ the Bethe approximation to compute the two-link marginals $\mu_{t\to t+1}^{ll'}(e_l,e_{l'})$. To this end, we introduce auxiliary variables $\delta l$ that allow us to know how the local changes affect on link $l$. More precisely, given $e_l(t)$ and $\delta l$ we will be able to recover $e_k(t-1)$ in the previous time step.  Here $\delta l$ takes a small number of values, as the number of possible local changes are small; the endpoints and the label of a link can at most change by $\pm 1$ in a LC-II.

Given the cavity marginals $\mu_{t\to t\pm 1}^{ll'}(e_l,e_{l'})$,  we obtain the local marginals $\mu_{t}^{ll'}(e_l,e_{l'})$ from $\mu_{t}(\mathsf{L}_t) \propto  e^{\beta E(t)}\mu_{t\to t+1}(\mathsf{L}_t) \mu_{t\to t-1}(\mathsf{L}_t)$. This allow us to find an estimation of the number of possible link configurations at time step $t$, as explained in Appendix \ref{ME-app}.
Figure \ref{f5} displays the information obtained in this way for different inverse temperatures $\beta$ with $M=10$ links. Finally,  to concentrate on the minimum evolutions, we take the limit $\beta \to \infty$ of the above equations to obtain the so called minsum equations \cite{KFL-inform-2001}, see Appendix \ref{ME-app}. These minsum equations are used in a reinforcement (smoothed decimation) algorithm \cite{BR-prl-2006} to find an approximate optimal path for given boundary conditions and time steps $T$.
      
\begin{figure}
\includegraphics[width=16cm]{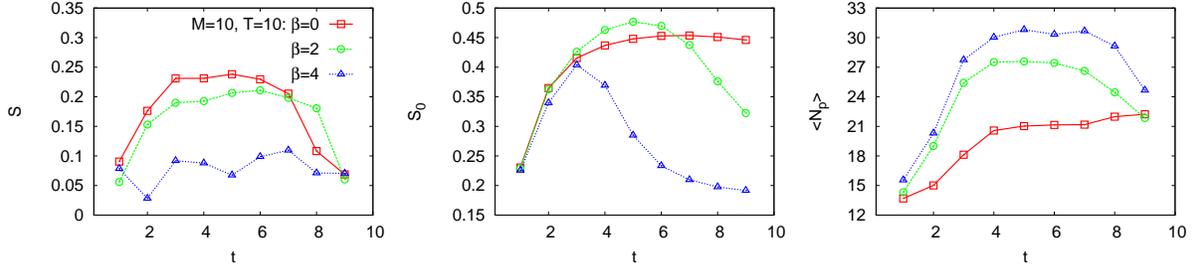} 
\caption{The entropy $S(t)$ in the intermediate steps of evolution from a random link configuration at $t=0$ to another random configuration at $t=10$ for $M=10$ links. The figure also shows $S_0(t)$, the entropy of configurations at distance $t$ from the initial configuration, and $\langle N_p(t) \rangle$ the average number of $N_p$ at time step $t$. The data have been obtained by the finite-temperature evolution algorithm with energy function $\mathcal{E}=- \sum_{t=1}^{T-1}N_p(t)$.}\label{f5}
\end{figure}

By the above approximate algorithm, we can explore the configuration space of larger number of links with larger evolution times. The time complexity of the algorithm grows like $TM^6$, and it takes still a few hours of CPU time to find an approximate minimum path of length $T=10$ for $M=10$. There are a few points here to mention about the algorithm. In each step we are approximating the cavity marginals $\mu_{t\to t\pm 1}(\mathsf{L})$ by a Bethe distribution that would degrade the algorithm performance as the evolution time $T$ increases. But, we are working with coarse-grained time steps allowing for $O(M)$ local changes of type II in each step. This is good because the average shortest distance between two randomly selected link configurations would grow as $M\ln M$ if the connectivity graph is close to a random graph of $\approx e^{M\ln M}$ nodes. It means that we do not need to work with very large number of coarse-grained time steps. This of course is obtained in the expense of optimizing a coarse-grained dynamics instead of the more detailed one.      
                   
\begin{figure}
\includegraphics[width=15cm]{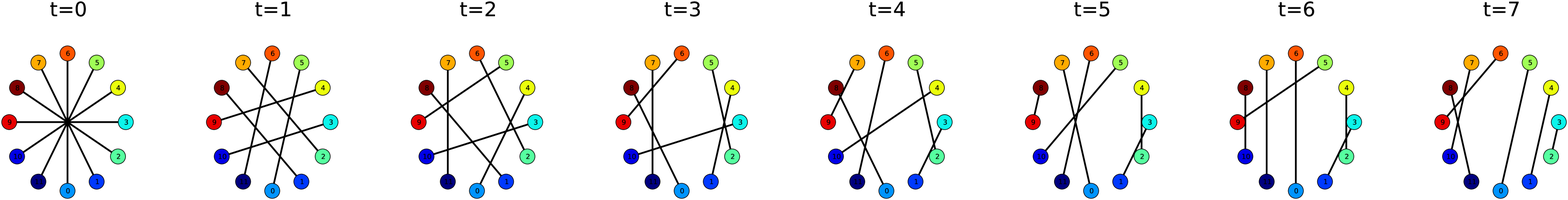} 
\includegraphics[width=15cm]{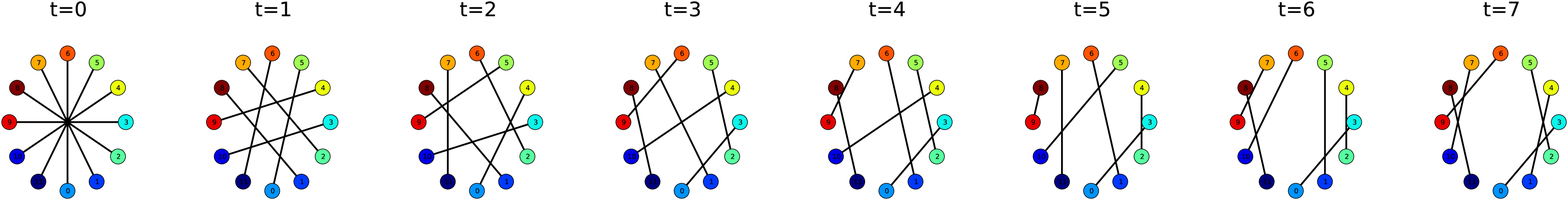} 
\caption{An evolution path of $M=6$ links for $T=7$ coarse-grained steps from the all-$x$ configuration to modular configurations of two components $p3x3$ and $x3x3$ obtained by the approximate minimum evolution algorithm minimizing $\mathcal{E}=-\sum_{t=1}^{T-1}N_p(t)$. The endpoints on the chain start from $i=0$ (at the bottom of the circle) and increase to $i=2M-1$ in the counter-clockwise direction.}\label{f6}
\end{figure}

As an example, we consider the evolution of $M=6$ links from the all-$x$ configuration $x6$ to a modular configuration of two components $p3x3$. The shortest path has length $T=12$ with $\mathcal{N}_p=30$. For an optimal path of length $T=16$ maximizing $\mathcal{N}_p$, we obtain $\mathcal{N}_p=62$ by the exact algorithm. On the other hand, using the approximate algorithm, we obtain a path of $7$ coarse-grained steps, shown in Fig. \ref{f6}, which can be decomposed into $T=19$ local changes of type II with $\mathcal{N}_p=54$. In the same figure, we show a path from $x6$ to $x3x3$. Here the shortest path takes $T=9$ steps and $\mathcal{N}_p=0$. For $T=13$ we find $\mathcal{N}_p=22$ by the exact algorithm whereas the approximate algorithm gives $\mathcal{N}_p=28$ in $7$ coarse-grained steps consisting of $15$ local changes of type II. Note that, however, these are not fair comparisons as the two algorithms are not optimizing exactly the same dynamics.  In Appendix \ref{figs-app}, we display more instances of evolutions obtained by the approximate algorithm for a larger system with $M=10$ links and $T=9$.

\section{Discussion}\label{S4}

Conformational transitions are ubiquitous in biomolecular systems, have significant functional role and are subject to evolutionary pressures. These transitions change properties of molecules including topology, surface properties and mechanical flexibility, and subsequently affect their affinity for interacting partners as well as functions. Despite existence of experimental data regarding the long-lived stable states of biomolecules, little or no experimental data are often available on the intermediate states along the conformational transition pathway associated with a function or evolution of a function. Theoretical and computational efforts are providing fundamental insights into this process. Here we provided a first theoretical framework for topological transition, i.e. conformational transitions that are associated with changes in molecular topology. 

We defined distance measures on the space of circuit topologies and then used the distance measures to study how topologies evolve under the most basic protocols. Topology of a chain can change in two ways, by changing the number of contacts or by rearranging the contacts \cite{MTM-pccp-2014}. Here we studied correlated disruption of two neighboring contacts and subsequent reformation of contacts via exchanging neighboring contact sites. This protocol is for example applicable to rearrangements due to co-variation of monomers that are close in primary sequence. It also assumes that two neighboring contacts have a higher chance of exchanging their partners due to physical proximity (closeness in sequence commonly leads to proximity in 3D). We also studied the dynamics under a protocol in which two random contacts are disrupted and the free contact sites form a new contact pair. In addition, we considered the dynamics with variable number of links, where a contact configuration can also change by creation or annihilation of a single contact. We defined connectivity graphs and distance measures based on these local changes in the space of contact configurations, and studied their statistical properties. %For example, maximizing the overlap of two configuration matrices resulted in distance $D_{III}\simeq 0.29 M(M-1)/2$ for two randomly generated link configurations.

Our analysis revealed properties of the topological space and their implications for topological dynamics. Using the exact algorithm for small number of links, we find: (1) The connectivity graph of link configurations that are related by the local changes of type II and $M^{\pm}$ shows that configurations are more concentrated around the modular configurations. (2) The optimal shortest paths  minimizing the energy functional $\mathcal{E}=-\sum_{t=1}^{T-1}N_p(t)$ behave very differently in the connectivity graphs obtained by the LC-II; we observed very large fluctuations in the degeneracy $g$ and energy gap $\Delta$, accompanied by nearly no correlation between these quantities and the path length. As another example, the optimal shortest paths in $\mathcal{G}_{I+M^{\pm}}$ exhibit in average smaller degeneracy and larger gap than the optimal shortest paths in $\mathcal{G}_{II+M^{\pm}}$.

For larger number of links, we devised an approximate algorithm to estimate the number of intermediate link configurations connecting two boundary configurations in a path of length $T$. A zero-temperature limit of the algorithm allows us to find an approximate optimal path minimizing an additive energy functional of the evolution. The challenge here is the study of larger problem sizes by more accurate and efficient approximations, to investigate large deviations (rare events) in the energy landscape of the evolution.   

The simplicity of our model allowed us to illustrate the notion of topological dynamics without the need to consider the complexity of a real system. Despite being a critical determinant of polymer dynamics \cite{MTM-pccp-2014,QM-macro-2014}, topology is often not the only factor governing molecular functions and dynamics. Many other factors such as chemical nature of the monomers, geometric constraints, solvent properties as well as interacting molecules play critical roles. Further experimental, bioinformatic and theoretical studies are needed to test the relevance, applicability and predictability of the proposed distance measures and modes of dynamics in real world applications. 

\section{Conclusion}

In this article, we explored how topology imposes constraints on chain dynamics and shapes the conformational search towards a desired state. The topology framework can be used for a host of applications in structural biology, evolutionary biology and materials sciences, among others. We conclude by suggesting possible directions for future research.

The topology framework is expected to find applications in analysis of genome architecture and dynamics. In particular, our approach can be readily applied to study the topologically associated domains (TADs) in chromosomes \cite{DSY-nature-2012,JCC-NAR-2012,PRDW-Nature-2014,ZW-pnas-2015}. Here, DNA acts as a polymer chain and binding proteins act as contacts bringing distinct sites on DNA together. Despite complexity of genomic architecture and nuclear environment, loci that are distant are often able to come to close proximity to form physical contacts \cite{SRB-PCB-2013} that subsequently lead to biochemical interactions and signalings \cite{PM-PRL-2013}. Contacts between genomic loci may break and re-establish into a new arrangement. The arrangement of these contacts and the frequencies of contact formation between sites on the same or different chromosomes can be measured using Chromosome Conformation Capture (C3) technology and its more recent variants such as  Hi-C techniques \cite{DDK-science-2002,BLW-JOVE-2010,DRA-GenRes-2006,ZTS-naturegenet-2006}. Various quantities analyzed in this article could assume a different meaning when put in this context: gap between, and degeneracy of, states would be related to the stability and frequency of 3D conformations. These could then directly inform biological experiments \cite{JCC-NAR-2012,AH-PRL-2013}. Dynamical changes between topological states could be interpreted as reactions catalyzed by histone modifications in the case of inter-phase chromosomes.

Our approach can also be used to study dynamics and evolution of proteins. As such one has to first extract circuit topology from coordinate files, as described in \cite{MWT-structure-2014}. One can then infer energy functions that serve to describe the statistical weight of contact configurations in a dynamical process. In Ref. \cite{MR-RSC-2015} we used local topological data to reconstruct such an energy function. There, we observed that appropriate forms of two-contact interactions are enough to describe simple structural orders like modules and sectors.

Characterizing the nature of large deviations of stochastic quantities (e.g. work) in a nonequilibrium process (e.g. stretching a polymer in contact with a thermal bath) is essential to our understanding of biophysical systems and in general the principles of nonequilibrium systems \cite{F-pnas-2003,CTW-JCP-2005,AFR-JCP-2006,MWZ-pre-2012,CC-RSC-2015}. Here one needs powerful and efficient sampling algorithms to capture the statistical properties of the rare events. The cavity method and algorithms we utilized in this work proved useful in the study of equilibrium glassy systems, and we expect to be helpful also in biophysical applications.             

Finally, we envision application of our approach to areas other than polymer physics. In particular, our study may be applicable to problems involving optimal evolution of dynamical (stochastic) systems.
Linearly ordered objects with inter-object interactions are ubiquitous in nature and in applications ranging from economics and operations research to biology. Understanding the linear order and the arrangements of inter-object interactions are crucial for understanding the functions and dynamics of these systems. The former issue, known as linear ordering problem or job shop scheduling, has been intensely studied over the last few decades. The latter, however, remained less understood. Here, we studied polymer chains as a prototype of linear chain of objects with intra-chain interactions in its topological space. We hope that our results will further stimulate research along these lines. 

\acknowledgments
We would like to thank R. Zecchina and anonymous referees for careful reading of the manuscript and their helpful comments.

\newpage
\appendix

\section{Details of the minimum assignment algorithm}\label{MD-app}
Consider two $M\times M$ matrices $\mathsf{A}^1$ and $\mathsf{A}^2$ with elements $\mathsf{A}_{ll'}^{1,2} \in \{p,s,x\}$ defining the relative position of links $l$ and $l'$. The aim is to find a one-to-one assignment $k(l)$ of the links from $\mathbf{l} \to \mathbf{k}$ minimizing the Hamming distance $D(\mathsf{A}^1,\mathsf{A}^2)=\sum_{l<l'} (1-\delta_{\mathsf{A}_{ll'}^1,\mathsf{A}_{kk'}^2})$. Here, to shorten the notation, we use $k,k'$ for $k(l),k(l')$.

We start from the local marginals $\mu_l(k)$ of the probability measure $\mu(\mathbf{k})\propto e^{-\beta D(\mathsf{A}^1,\mathsf{A}^2)}$ of the assignments written in the Bethe approximation for a finite $\beta$ \cite{MM-book-2009},
\begin{align}
\mu_{l}(k) \propto \prod_{l'\neq l}\left( \sum_{k' \neq k} e^{-\beta (1-\delta_{\mathsf{A}_{ll'}^1,\mathsf{A}_{kk'}^2})} \mu_{l'\to l}(k')\right).
\end{align}
The cavity marginals $\mu_{l'\to l}(k')$ give the probability of assigning $l' \to k'$ in the absence of link $l$. The recursive equations governing these cavity marginals are called the belief propagation equations \cite{MM-book-2009,KFL-inform-2001}, 
\begin{align}
\mu_{l'\to l}(k') \propto \prod_{l''\neq l,l'}\left( \sum_{k'' \neq k'} e^{-\beta (1-\delta_{\mathsf{A}_{l'l''}^1,\mathsf{A}_{k'k''}^2})} \mu_{l''\to l'}(k'')\right).
\end{align}
We can solve the equations by iteration starting from a random initial condition.

But we are interested in the limit $\beta \to \infty$ of the equations concentrating on the optimal assignments minimizing $D(\mathsf{A}^1,\mathsf{A}^2)$. Assuming the scaling $\mu_{l\to l'}(k)=e^{-\beta h_{l\to l'}(k)}$ for the cavity marginals, the limit $\beta \to \infty$ of the BP equations read
\begin{align}
h_{l'\to l}(k')= \sum_{l''\neq l,l'} \min_{k'' \neq k'} \left\{ (1-\delta_{\mathsf{A}_{l'l''}^1,\mathsf{A}_{k'k''}^2}) + h_{l''\to l'}(k'')\right\}-C_{l'\to l}.
\end{align}
Here $C_{l'\to l}$ is a constant to make $\min_{k'} h_{l'\to l}(k')=0$. These equations are called minsum equations \cite{KFL-inform-2001}. 

We use the above equations in a reinforcement algorithm to fix smoothly the assignment variables \cite{BR-prl-2006}. To this end, we use the information in the local marginals $\mu_{l}(k)=e^{-\beta h_{l}(k)}$ to increase slowly an external field acting on the variables. The aim is to concentrate more and more the cavity and local marginals on a minimum assignment as the algorithm proceeds. More precisely, we start from random initial messages $h_{l}^0(k),h_{l \to l'}^0(k)$, and in each step we update the message in the following way:
\begin{align}
h_{l\to l'}^{t+1}(k)= \eta_l(k) + r(t) h_{l}^{t}(k)+\sum_{l''\neq l,l'} \min_{k'' \neq k} \left\{ (1-\delta_{\mathsf{A}_{ll''}^1,\mathsf{A}_{kk''}^2}) + h_{l''\to l}^t(k'')\right\}-C_{l\to l'}.
\end{align}
In the same way, we update the local messages
\begin{align}
h_{l}^{t+1}(k)= \eta_l(k) +r(t) h_{l}^{t}(k)+\sum_{l''\neq l} \min_{k'' \neq k} \left\{ (1-\delta_{\mathsf{A}_{ll''}^1,\mathsf{A}_{kk''}^2}) + h_{l''\to l}^t(k'')\right\}-C_{l}.
\end{align}
Here $r(t)$ is the reinforcement parameter; it is zero at the beginning and increases slowly by time as $r(t+1)=r(t)+\delta r$, for a small $\delta r \simeq 0.01$.  
In addition, we introduced a small noise $\eta_l(k)$ to the equations to reduce the number of possible minimum assignments.  
In each iteration one updates all the local and cavity messages selected in a random sequential way, according to the above equations. 
In the end, one obtains an assignment by looking at the local messages; that is $l \to k=\arg\min h_l^t(k)$.

\section{Ordering statistics of the local changes}\label{ordering-app}
Consider two link configurations $\mathsf{L}_0,\mathsf{L}_T$ connected by a sequence of local changes, that is $\mathsf{L}_T=\mathbf{u}_T\cdots \mathbf{u}_1\mathsf{L}_0$. To be specific, we assume the $\mathbf{u}$s are local changes (LC) of type II, where $\mathbf{u}$ is an elementary permutation of the neighboring endpoints $(i_u,i_u+1)$. There could be different orderings of the LCs connecting the same boundary configurations. The question is how these different orderings affect on a macroscopic behavior (phenotype) of the chain, for example a monotonically increasing (fitness) function of the $N_{p,s,x}$.   

For simplicity, let us ignore the link labels and work with the connectivity patterns of the endpoints $\mathsf{C}$. We will also focus on the simple case of two local changes $\mathbf{u},\mathbf{v}$ of type II. The local change $\mathbf{u}$ commutes with $\mathbf{v}$ in the context of $\mathsf{C}$ if $\mathbf{v}\mathbf{u}\mathsf{C}=\mathbf{u}\mathbf{v}\mathsf{C}$. Note that the transformations are reversible, that is from $\mathsf{C}'=\mathbf{v}\mathbf{u}\mathsf{C}$ we obtain $\mathsf{C}=\mathbf{u}\mathbf{v}\mathsf{C}'$. The definitions can readily be extended to link configurations with distinguishable links as well.

The two local changes $\mathbf{u},\mathbf{v}$ may involve two, three, or four distinct links; we will not consider permutation of neighboring endpoints that belong to a single link, because it has no effect. In table \ref{table-1} we summarize the possible effects of two commutative local changes on a contact configuration, considering only the nontrivial case of three links. One can easily construct the other cases with two or four links, following the above rules. Note that each transformation in the table can also happen in the reverse direction. As the table shows, the changes in numbers $N_{p,s,x}$ in the two paths are correlated depending on how that quantity changes form $\mathsf{C}$ to $\mathsf{C}'$. In particular, when $N_{q}$ increases (or decreases) the corresponding changes in the two paths do not have different signs; a LC-II only changes the state of two links from x to (p,s), or from (p,s) to x. This means that in a LC-II we have $\delta N_{p,s}=-\delta N_x=\pm 1$, and two LC-II can at most change $N_{p,s,x}$ by two. Consequently,  if $N_q$ increases (decreases), the changes $\delta N_q$ resulted by the two LC-II can not have different signs, because they can not give the expected total variation in $N_q$.

\begin{table}
\begin{center}

\begin{tabular}{|c|c|c|c||c|c|c|c|}
  \hline

    $p^{N_p}s^{N_s}x^{N_x}$     & $N_p$ &  $N_s$ & $N_x$ &  $p^{N_p}s^{N_s}x^{N_x}$ & $N_p$ & $N_s$ & $N_x$ \\

  \hline

    $p^3\to (p^2x,p^2x) \to px^2$     & $(\downarrow,\downarrow)$ &  $(-,-)$ & $(\uparrow,\uparrow)$ & 

    $s^3\to (s^2x,s^2x) \to sx^2$     & $(-,-)$ &  $(\downarrow,\downarrow)$ & $(\uparrow,\uparrow)$  \\

  \hline

    $x^3\to (px^2,sx^2) \to psx$     & $(\uparrow,-)$ &  $(-,\uparrow)$ & $(\downarrow,\downarrow)$ &  

    $x^3\to (px^2,px^2) \to p^2x$     & $(\uparrow,\uparrow)$ &  $(-,-)$ & $(\downarrow,\downarrow)$  \\

  \hline
  
    $ps^2\to (psx,s^2x) \to sx^2$     & $(-,\downarrow)$ &  $(\downarrow,-)$ & $(\uparrow,\uparrow)$ & 
       
    $px^2\to (p^2x,psx)\to p^2s$     & $(\uparrow,-)$ &  $(-,\uparrow)$ & $(\downarrow,\downarrow)$   \\

  \hline

    $px^2\to (psx,x^3)\to sx^2$     & $(-,\downarrow)$ &  $(\uparrow,-)$ & $(\downarrow,\uparrow)$ & 

    $px^2\to (p^2x,x^3)\to px^2$     & $(\uparrow,\downarrow)$ &  $(-,-)$ & $(\downarrow,\uparrow)$  \\

  \hline

    $p^2s\to (psx,psx)\to sx^2$     & $(\downarrow,\downarrow)$ &  $(-,-)$ & $(\uparrow,\uparrow)$ &  

    $p^2s\to (psx,p^2x)\to px^2$     & $(\downarrow,-)$ &  $(-,\downarrow)$ & $(\uparrow,\uparrow)$   \\

  \hline

    $p^2x\to (px^2,p^2s)\to psx$     & $(\downarrow,-)$ &  $(-,\uparrow)$ & $(\uparrow,\downarrow)$ &  

    $p^2x\to (px^2,p^3)\to p^2x$     & $(\downarrow,\uparrow)$ &  $(-,-)$ & $(\uparrow,\downarrow)$   \\

  \hline

    $s^2x\to (ps^2,sx^2)\to psx$     & $(\uparrow,-)$ &  $(-,\downarrow)$ & $(\downarrow,\uparrow)$ &  

    $s^2x\to (s^3,sx^2)\to s^2x$     & $(-,-)$ &  $(\uparrow,\downarrow)$ & $(\downarrow,\uparrow)$   \\

  \hline

    $psx\to (sx^2,p^2s)\to psx$     & $(\downarrow,\uparrow)$ &  $(-,-)$ & $(\uparrow,\downarrow)$ &  

    $$     & $$ &  $$ & $$   \\

  \hline

\end{tabular}

\vskip 0.5cm

\caption{The set of distinct transformations $\mathsf{C}\to (\mathbf{u}\mathsf{C},\mathbf{v}\mathsf{C})\to \mathsf{C}'= \mathbf{u}\mathbf{v}\mathsf{C}=\mathbf{v}\mathbf{u}\mathsf{C}$ obtained by two commutative local changes of type II applied on three links. The arrows in each column show the change in the numbers $N_{p,s,x}$: positive ($\uparrow$) or negative ($\downarrow$). The first (second) arrow in the parenthesis corresponds to the first (second) transition. Here $p^{N_p}s^{N_s}x^{N_x}$ shows a configuration of $N_q$ contact pairs of type $q=p,s,x$.}\label{table-1}
\end{center}
\end{table}

\section{Details of the minimum evolution algorithm}\label{ME-app}
Let us start from the dynamical partition function
\begin{align}
\mathcal{Z}(\mathsf{L}_0\to \mathsf{L}_T)=\sum_{\mathbf{u}(1),\mathbf{u}(2),\dots, \mathbf{u}(T)} e^{-\beta \mathcal{E}} \delta_{\mathsf{L}(\mathbf{u}(1),\mathbf{u}(2),\dots,\mathbf{u}(T)|\mathsf{L}_0),\mathsf{L}_T},
\end{align}
where $\mathsf{L}_t=\mathsf{L}(\mathbf{u}(1),\mathbf{u}(2),\dots,\mathbf{u}(t)|\mathsf{L}_0)$ is the link configuration at time step $t$, and $\mathbf{u}(t)$ defines the position of the possible local changes. In the following, we assume the local changes are of type II.     
Here $\mathcal{E}=\sum_{t=1}^{T-1}E(t)+ \sum_{t=1}^{T}E(t-1,t)$ with $E(t)=-N_p(t)$, and $E(t-1,t)=-\sum_{q<q'}\lambda_{q\to q'}N_{q\to q'}(t-1,t)$ for $q=p,s,x$. We recall that a link configuration is defined by the endpoints of all the links, and any two links have different endpoints. A local-change configuration $\mathbf{u}(t)=\{u_{ll'}(t)=0,1|\sum_{l'\ne l}u_{ll'}(t)\le 1\}$ is a matching of neighboring links with adjacent endpoints along the contact chain.  
      
We solve the above problem by a dynamic programming (message-passing) algorithm: Define the cavity messages $\mu_{t\to t+1}(\mathsf{L}_t)$ and $\mu_{t\to t-1}(\mathsf{L}_t)$ as the probability of having link configuration $\mathsf{L}_t$ in the absence of the energy terms and constraints imposed by the other part of the system; i.e. the segment $(t,T]$ for the forward message $\mu_{t\to t+1}$, and $[0,t)$ for the backward message $\mu_{t\to t-1}$. From the above partition function we can easily write the equations for these cavity marginals
\begin{align}
\mu_{t\to t+1}(\mathsf{L}_t) &=\frac{1}{z_{t\to t+1}} e^{-\beta E(t)}\sum_{\mathbf{u}(t)} \delta_{\mathsf{L}(\mathbf{u}(t)|\mathsf{L}_{t-1}),\mathsf{L}_t} e^{-\beta E(t-1,t)}\mu_{t-1\to t}(\mathsf{L}_{t-1}),\\
\mu_{t\to t-1}(\mathsf{L}_t) &=\frac{1}{z_{t\to t-1}} e^{-\beta E(t)}\sum_{\mathbf{u}(t+1)} \delta_{\mathsf{L}(\mathbf{u}(t+1)|\mathsf{L}_t),\mathsf{L}_{t+1}} e^{-\beta E(t,t+1)}\mu_{t+1\to t}(\mathsf{L}_{t+1}).  
\end{align}
Then the total marginal at time step $t$ is given by
\begin{align}
\mu_{t}(\mathsf{L}_t)=\frac{1}{z_t} e^{\beta E(t)}\mu_{t\to t+1}(\mathsf{L}_t) \mu_{t\to t-1}(\mathsf{L}_t).  
\end{align}  
The $z_{t\to t\pm 1}$ and $z_t$ are normalization constants.

\subsection{Approximating the messages}
We represent a link configuration $\mathsf{L}$ by the set of endpoints $e_l=(i_l,j_l)$, and label the links according to the order of their first endpoints. We also approximate the cavity messages by a Bethe distribution
\begin{align}
\mu_{t-1\to t}(\mathsf{L}) \approx \prod_l \mu_{t-1\to t}^l(e_l) \prod_{l<l'} \frac{\mu_{t-1\to t}^{ll'}(e_l,e_{l'})}{\mu_{t-1\to t}^l(e_l)\mu_{t-1\to t}^{l'}(e_{l'})}.
\end{align}
Using this structure for the cavity messages in the right hand side of the equations, we obtain the equations for the two-link marginals $\mu_{t\to t+1}^{ll'}(e_l,e_{l'})$, 
\begin{multline}
\mu_{t\to t+1}^{ll'}(e_l(t),e_{l'}(t)) \propto \sum_{\{e_{l''}(t)|l''\ne l,l'\}} e^{-\beta E(t)}\sum_{\mathbf{u}(t)} \delta_{\mathsf{L}(\mathbf{u}(t)|\mathsf{L}_{t-1}),\mathsf{L}_t} e^{-\beta E(t-1,t)} \\ \times \prod_k \mu_{t-1\to t}^k(e_k(t-1)) \prod_{k<k'} \frac{\mu_{t-1\to t}^{kk'}(e_k(t-1),e_{k'}(t-1))}{\mu_{t-1\to t}^k(e_k(t-1))\mu_{t-1\to t}^{k'}(e_{k'}(t-1))}.
\end{multline}
    
We compute the sum in the right hand side of the above equation using the Bethe approximation \cite{MM-book-2009}. To this end, we introduce auxiliary variables $\delta l$ to see how the local changes affect link $l$. More precisely, given $e_l(t)$ and $\delta l$ we can recover the endpoints and the link label $e_k(t-1)$ in previous step.  Note that $\delta l$ takes a small number of values as the number of possible local changes of type II are small; the endpoints and the label of a link can at most change by $\pm 1$. The approximate two-link marginal $\mu_{t\to t+1}^{ll'}(e_l,e_{l'})$ can be obtained by considering the constraints involving $(e_l,e_{l'})$, and by taking into account the effect of the remaining degrees of freedom. The latter is provided by a new set of cavity marginals $\nu_{l\to l'}(e_l(t);\delta l;u_{ll'}(t))$ giving the probability of indicated variables in the absence of $l'$. Putting all together, we obtain 
\begin{multline}\label{cll-app-1}
\mu_{t\to t+1}^{ll'}(e_l(t),e_{l'}(t)) \propto  \sum_{\delta l,\delta l',u_{ll'}(t)}w_{ll'}(e_l(t),e_{l'}(t);\delta l,\delta l';u_{ll'}(t)) \\ \times \nu_{l\to l'}(e_l(t);\delta l;u_{ll'}(t))\nu_{l'\to l}(e_{l'}(t);\delta l';u_{ll'}(t)),
\end{multline}
where we defined
\begin{align}\label{cll-app-2}
w_{ll'}(e_l(t),e_{l'}(t);\delta l,\delta l';u_{ll'}(t))=w_{l'\to l}(e_l(t),e_{l'}(t);\delta l,\delta l';u_{ll'}(t))\mu_{t-1\to t}^k(e_k(t-1)),
\end{align}
with 
\begin{multline}\label{cll-app-3}
w_{l'\to l}(e_l(t),e_{l'}(t);\delta l,\delta l';u_{ll'}(t))= \mathbb{I}(\delta l,\delta l',u_{ll'}(t)|e_l(t),e_{l'}(t))  e^{\beta \delta_{\mathsf{q}(e_l(t),e_{l'}(t)),p}} \\ \times e^{u_{ll'}(t) \beta \lambda_{\mathsf{q}(e_k(t-1),e_{k'}(t-1))\to \mathsf{q}(e_l(t),e_{l'}(t))}}
\mu_{t-1\to t}^{kk'}(e_{k'}(t-1)|e_k(t-1)).
\end{multline}
Here $\mathbb{I}(\delta l,\delta l',u_{ll'}(t)|e_l(t),e_{l'}(t))$ is an indicator function to ensure that: (i) $e_l(t) \neq e_{l'}(t)$, (ii) the links are labeled from  left to right according to their first endpoints, and (iii) to check for the possibility of a local change given the endpoints and the $\delta l,\delta l',u_{ll'}(t)$. Moreover, $\mu_{t-1\to t}^{kk'}(e_{k'}(t-1)|e_k(t-1))=\mu_{t-1\to t}^{kk'}(e_{k'}(t-1),e_k(t-1))/\mu_{t-1\to t}^{k}(e_k(t-1))$ is the conditional probability of $e_{k'}(t-1)$ given $e_{k}(t-1)$, and $\mathsf{q}(e_l,e_{l'}) \in \{p,s,x\}$ depending on the link endpoints.

The $\nu_{l\to l'}(e_l(t);\delta l;u_{ll'}(t))$ are determined by the following Bethe equations:
\begin{multline}\label{cll-app-4}
\nu_{l\to l'}(e_l(t);\delta l;0) \propto  \prod_{l'' \ne l,l'}\left( \sum_{e_{l''}(t),\delta l''} w_{l''\to l}(e_l(t),e_{l''}(t);\delta l,\delta l'';0) \nu_{l'' \to l}(e_{l''}(t);\delta l'';0) \right) \\+ \sum_{l'' \ne l,l'}\left( \sum_{e_{l''}(t),\delta l''} w_{l''\to l}(e_l(t),e_{l''}(t);\delta l,\delta l'';1) \nu_{l'' \to l}(e_{l''}(t);\delta l'';1) \right)\\ \times
\prod_{l''' \ne l,l',l''}\left( \sum_{e_{l'''}(t),\delta l'''} w_{l'''\to l}(e_l(t),e_{l'''}(t);\delta l,\delta l''';0) \nu_{l''' \to l}(e_{l'''}(t);\delta l''';0) \right),
\end{multline}
and,
\begin{align}\label{cll-app-5}
\nu_{l\to l'}(e_l(t);\delta l;1) \propto  \prod_{l'' \ne l,l'}\left( \sum_{e_{l''}(t),\delta l''} w_{l''\to l}(e_l(t),e_{l''}(t);\delta l,\delta l'';0) \nu_{l'' \to l}(e_{l''}(t);\delta l'';0) \right). 
\end{align}

Similarly, we obtain the cavity marginals $\mu_{t\to t-1}^{ll'}(e_l(t),e_{l'}(t))$, and finally the local marginals read
\begin{align}\label{mll-app-1}
\mu_t^{ll'}(e_l(t),e_{l'}(t)) \propto  \nu_{l\to l'}(e_l(t)) \tilde{w}_{ll'}(e_l(t),e_{l'}(t)) \nu_{l'\to l}(e_{l'}(t)), 
\end{align}
where now
\begin{align}\label{mll-app-2}
\tilde{w}_{ll'}(e_l(t),e_{l'}(t))= \tilde{w}_{l'\to l}(e_l(t),e_{l'}(t)) \mu_{t\to t+1}^{l}(e_l(t))\mu_{t\to t-1}^{l}(e_l(t)),
\end{align}
with
\begin{align}\label{mll-app-3}
\tilde{w}_{l'\to l}(e_l(t),e_{l'}(t))=  e^{-\beta \delta_{\mathsf{q}(e_l(t),e_{l'}(t)),p}} \mu_{t\to t+1}^{ll'}(e_{l'}(t)|e_l(t))\mu_{t\to t-1}^{ll'}(e_{l'}(t)|e_l(t)),
\end{align}
and, 
\begin{align}\label{mll-app-4}
\nu_{l\to l'}(e_l(t)) \propto \prod_{l'' \neq l,l'}\left( \sum_{e_{l''}(t)} \tilde{w}_{l''\to l}(e_l(t),e_{l''}(t))  \nu_{l''\to l}(e_{l''}(t)) \right). 
\end{align}

Given the local marginals $\mu_t^{l}(e_l(t))$ and $\mu_t^{ll'}(e_l(t),e_{l'}(t))$, an estimation of the entropy at time step $t$ can be obtained by the Bethe entropy \cite{MM-book-2009},
\begin{align}\label{s-app-1}
S(t)=\frac{1}{M\ln M}\left(\sum_{l<l'}\Delta S_{ll'}-(M-2)\sum_{l=1}^M\Delta S_{l}\right), 
\end{align}
where 
\begin{align}\label{s-app-2}
\Delta S_l &=-\sum_{e_l} \mu_t^{l}(e_l) \ln \mu_t^{l}(e_l),\\
\Delta S_{ll'} &=-\sum_{e_l,e_{l'}} \mu_t^{ll'}(e_l,e_{l'}) \ln \mu_t^{ll'}(e_l,e_{l'}).
\end{align}

In summary, from Eq. \ref{cll-app-1} we obtain the cavity marginals $\mu_{t\to t+1}^{ll'}(e_l(t),e_{l'}(t))$, and similarly for $\mu_{t\to t-1}^{ll'}(e_l(t),e_{l'}(t))$. Then, Eq. \ref{mll-app-1} gives the local marginals $\mu_{t}^{ll'}(e_l(t),e_{l'}(t))$, which are used in Eq. \ref{s-app-1} to compute the Bethe entropy.

\subsection{The zero temperature limit $\beta \to \infty$}
To take the limit $\beta \to \infty$, we assume the above probability distributions scale as
\begin{align}
\mu_{t\to t+1}^{l}(e_l(t)) &=e^{-\beta h_{t\to t+1}^l(e_l(t))},\\
\mu_{t\to t+1}^{ll'}(e_l(t),e_{l'}(t)) &=e^{-\beta h_{t\to t+1}^{ll'}(e_l(t),e_{l'}(t))},
\end{align}
and similarly for the messages from $t$ to $t-1$. In addition, we define 
\begin{align}
\nu_{l\to l'}(e_l(t);\delta l;u_{ll'}(t)) &= e^{-\beta g_{l\to l'}(e_l(t);\delta l;u_{ll'}(t))},\\
\nu_{l\to l'}(e_l(t)) &= e^{-\beta g_{l\to l'}(e_l(t))}.
\end{align}

Now the zero temperature (minsum) equations read \cite{MM-book-2009,KFL-inform-2001},
\begin{multline}\label{cms-app-1}
h_{t\to t+1}^{ll'}(e_l(t),e_{l'}(t)) =  \min_{\delta l,\delta l',u_{ll'}(t): \mathbb{I}} \Big\{ -u_{ll'}(t)\lambda_{\mathsf{q}(e_k(t-1),e_{k'}(t-1))\to \mathsf{q}(e_l(t),e_{l'}(t))}  \\-\delta_{\mathsf{q}(e_l(t),e_{l'}(t)),p}+ h_{t-1\to t}^{kk'}(e_k(t-1),e_{k'}(t-1)) \\+ g_{l\to l'}(e_l(t);\delta l;u_{ll'}(t))+ g_{l'\to l}(e_{l'}(t);\delta l';u_{ll'}(t)) \Big\},
\end{multline}
where the minimum is subject to the constraints in $\mathbb{I}(\delta l,\delta l',u_{ll'}(t)|e_l(t),e_{l'}(t))$, and
\begin{align}\label{cms-app-2}
g_{l\to l'}(e_l(t);\delta l;0) &=  \min\big\{F_{l\to l'}^0(e_l(t),\delta l),\min_{l''\neq l,l'} F_{l\to l'}^{l''}(e_l(t),\delta l)\big\},\\
g_{l\to l'}(e_l(t);\delta l;1) &=  F_{l\to l'}^0(e_l(t),\delta l).
\end{align}
Here we defined
\begin{align}\label{cms-app-3}
F_{l\to l'}^0(e_l(t),\delta l) &= \sum_{l'' \ne l,l'}f_{l''\to l}^0(e_l(t),\delta l),\\
F_{l\to l'}^{l''}(e_l(t),\delta l) &= f_{l''\to l}^1(e_l(t),\delta l)+\sum_{l''' \ne l,l',l''}f_{l'''\to l}^0(e_l(t),\delta l),\\
\end{align}
with
\begin{multline}\label{cms-app-4}
f_{l''\to l}^0(e_l(t),\delta l) = \min_{e_{l''}(t),\delta l'': \mathbb{I},u_{ll''}(t)=0} \Big\{-\delta_{\mathsf{q}(e_l(t),e_{l''}(t)),p} \\  + h_{t-1\to t}^{kk''}(e_k(t-1),e_{k''}(t-1)) 
-h_{t-1\to t}^{k}(e_k(t-1)) +g_{l'' \to l}(e_{l''}(t);\delta l'';0) \Big\},
\end{multline}
and
\begin{multline}\label{cms-app-5}
f_{l''\to l}^1(e_l(t),\delta l) =\min_{e_{l''}(t),\delta l'':\mathbb{I},u_{ll''}(t)=1} \Big\{ -\lambda_{\mathsf{q}(e_k(t-1),e_{k''}(t-1))\to \mathsf{q}(e_l(t),e_{l''}(t))}  -\delta_{\mathsf{q}(e_l(t),e_{l''}(t)),p}\\ + h_{t-1\to t}^{kk''}(e_k(t-1),e_{k''}(t-1)) 
-h_{t-1\to t}^{k}(e_k(t-1)) + g_{l'' \to l}(e_{l''}(t);\delta l'';1) \Big\}.
\end{multline}

Similarly we obtain the minsum messages $h_{t\to t-1}^{ll'}(e_l(t),e_{l'}(t))$, and finally the local messages read
\begin{multline}\label{ms-app-1}
h_t^{ll'}(e_l(t),e_{l'}(t))= \delta_{\mathsf{q}(e_l(t),e_{l'}(t)),p} + h_{t\to t+1}^{ll'}(e_{l'}(t),e_l(t))+h_{t\to t-1}^{ll'}(e_{l'}(t),e_l(t)) \\ +g_{l\to l'}(e_l(t))+ g_{l'\to l}(e_{l'}(t)), 
\end{multline}
and, 
\begin{multline}\label{ms-app-2}
g_{l\to l'}(e_l(t))= \sum_{l'' \neq l,l'} \min_{e_{l''}(t)\neq e_l(t)}\Big\{ \delta_{\mathsf{q}(e_l(t),e_{l''}(t)),p} + h_{t\to t+1}^{ll''}(e_{l''}(t),e_l(t))\\-h_{t\to t+1}^{l}(e_l(t))+h_{t\to t-1}^{ll''}(e_{l''}(t),e_l(t))-h_{t\to t-1}^{l}(e_l(t)) + g_{l''\to l}(e_{l''}(t)) \Big\}. 
\end{multline}

We use the above equations in a reinforcement algorithm to find a minimum evolution path satisfying all the connectivity constraints.    
In a reinforcement algorithm, we use the information in the local messages $h_t^{ll'}(e_l(t),e_{l'}(t))$ to slowly polarize the cavity messages $h_{t\to t\pm 1}^{ll'}(e_l(t),e_{l'}(t))$  
in the direction favored by the local messages, as we did in Appendix \ref{MD-app} for the minimum distance algorithm.    

In summary, the cavity messages $h_{t\to t+1}^{ll'}(e_l(t),e_{l'}(t))$ are obtained by solving Eq. \ref{cms-app-1} (similarly for $h_{t\to t-1}^{ll'}(e_l(t),e_{l'}(t))$). These messages are used in Eq. \ref{ms-app-1} to compute the local messages $h_t^{ll'}(e_l(t),e_{l'}(t))$ which are utilized in a reinforcement algorithm to find an approximate optimal pathway. 

\section{More details and figures}\label{figs-app}
In this section, we give more details of the numerical data and figures obtained in this study.

\begin{figure}
\includegraphics[width=10cm]{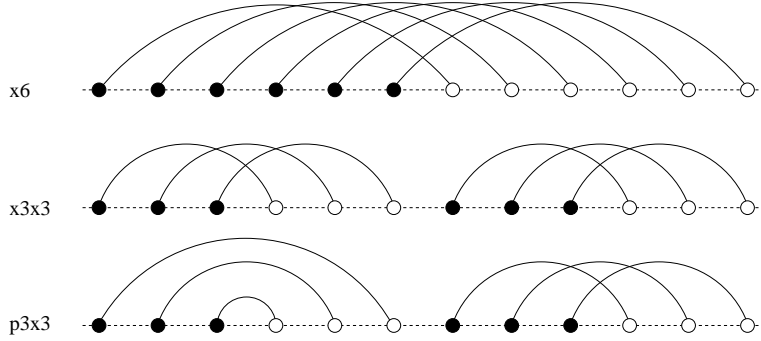} 
\caption{Examples of contact configurations of $M=6$ links used as boundary configurations in the minimum evolution algorithm: $x6$ (top), $x3x3$(middle), and $p3x3$ (bottom).}\label{f0-app}
\end{figure}

We are interested in topological evolutions connecting two boundary contact configurations $(\mathsf{L}_0,\mathsf{L}_T)$ by a sequence of local changes in the link arrangements. More specifically, we look for optimal pathways of length $T$  minimizing the energy functional $\mathcal{E}=-\sum_{t=1}^{T-1}N_p(t)\equiv -\mathcal{N}_p$, or $\mathcal{E}=- \sum_{t=1}^{T}[N_{x\to p}(t-1,t)+N_{x\to s}(t-1,t)]\equiv -\mathcal{N}_{x\to p,s}$.
Here $\mathcal{N}_p$ is the total number of contact pairs of type $p$, and $\mathcal{N}_{x\to p,s}$ gives the total number of contact pairs changed from $x$ to $p,s$ during the evolution. Figure \ref{f0-app} shows some boundary contact configurations we use in the following examples. For now, we assume the paths are simple with no loops, that is each configuration in the path is visited only once. 

\begin{figure}
\includegraphics[width=16cm]{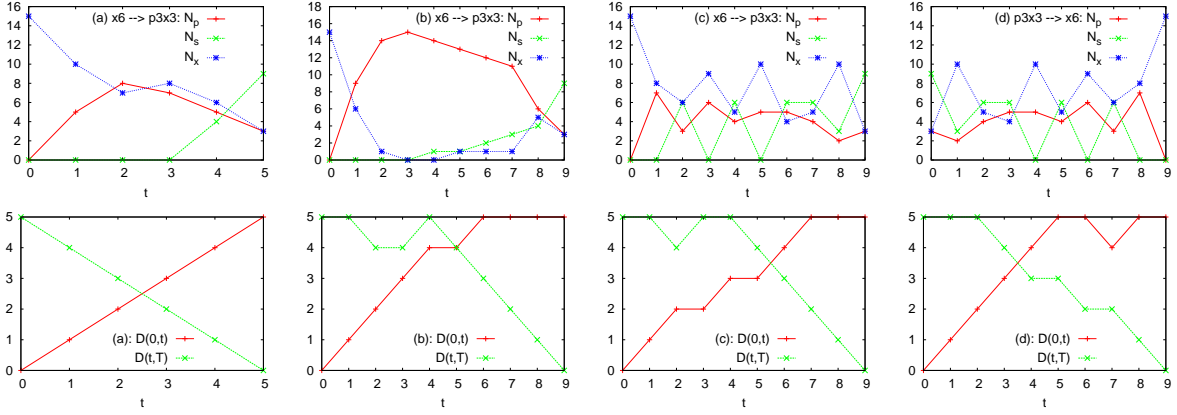} 
\caption{Evolution with local changes of type I: (top) $N_{p,s,x}(t)$, and (bottom) $D_I(t)$ of the intermediate link configurations from the boundary configurations in the paths obtained by the exact algorithm for $M=6$ links. The paths connect the all-$x$ ($x6$) configuration to a modular structure of two components ($p3x3$) at shortest distance $D_{I}=5$. Besides the shortest path (a), we display the path maximizing $\mathcal{N}_p$ (b), and the path maximizing $\mathcal{N}_{x\to p,s}$ for $x6\to p3x3$ (c) and $p3x3\to x6$ (d). Here $t$ denotes the number of local changes of type I. The path degeneracy $g$ and energy gap $\Delta$ are: $(g=2,\Delta=2)_a$, $(g=8,\Delta=2)_b$, $(g=1,\Delta=4)_c$, $(g=1,\Delta=4)_d$.}\label{f1-app}
\end{figure}

\begin{figure}
\includegraphics[width=12cm]{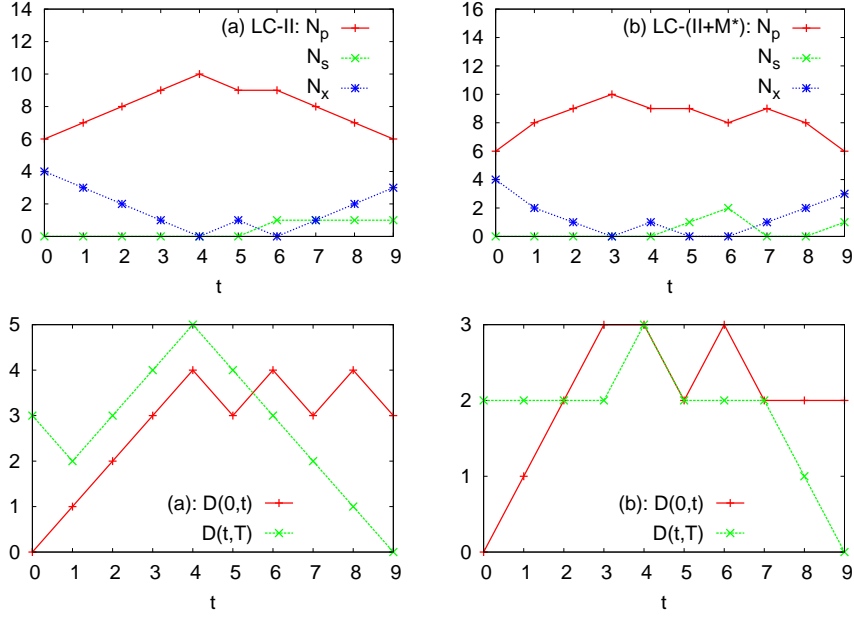} 
\caption{Evolution with local changes of type II (a,b) and (II+$M^*$) (c,d): (top) $N_{p,s,x}(t)$, and (bottom) distance $D(t)$ of the intermediate link configurations from two random boundary configurations of $M=5$ links. The boundary configurations have shortest distance $D_{II}=3$, and the optimal paths are obtained by the exact algorithm maximizing $\mathcal{N}_p$. Besides the optimal shortest paths (a,c), we display the results for a larger evolution time $T=9$ (b,d). Here $t$ denotes the number of local changes of type II (a,b) or II+$M^*$ (c,d). The path degeneracy $g$ and energy gap $\Delta$ are: $(g=1,\Delta=14)_a$, $(g=78,\Delta=1)_b$, $(g=1,\Delta=2)_c$, $(g=64,\Delta=1)_d$.}\label{f2-app}
\end{figure}

\begin{figure}
\includegraphics[width=12cm]{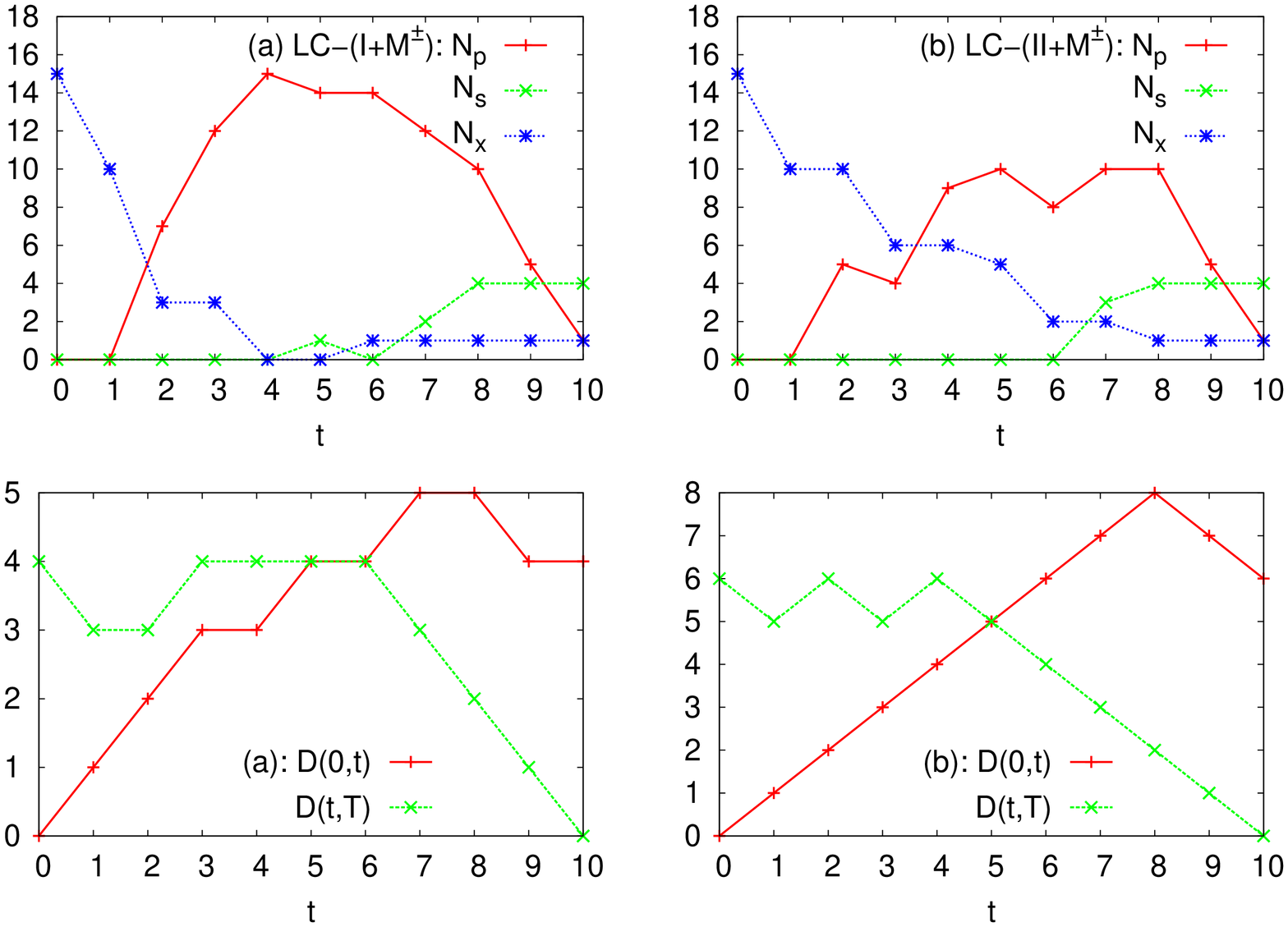} 
\caption{Evolution with local changes of type I+$M^{\pm}$ (a) and II+$M^{\pm}$ (b): (top) $N_{p,s,x}(t)$, and (bottom) distance $D(t)$ of the intermediate link configurations from the boundary configurations for $T=10$ steps. The boundary configurations $(x6,p2x2)$ have different number of links with shortest distances $4$(a) and $6$(b). The optimal paths are obtained by the exact algorithm maximizing $\mathcal{N}_p$. Here $t$ denotes the number of local changes of type I+$M^{\pm}$ (a) or II+$M^{\pm}$ (b). The path degeneracy $g$ and energy gap $\Delta$ are: $(g=1,\Delta=1)_a$, $(g=3,\Delta=1)_b$.}\label{f3-app}
\end{figure}

Figure \ref{f1-app} shows the optimal pathways from the all-$x$ ($x6$) configuration of $M=6$ links to the modular structure $p3x3$, following the local changes of type I. The results have been obtained by an exact algorithm searching in the space of all paths connecting the two boundary configurations. 
In Fig. \ref{f2-app}, we compare the optimal paths connecting two random link configurations of $M=5$ links with local changes of type II and (II+$M^*$). In the latter case, two link configurations are connected if they are related either by LC-II or LC-$M^*$. Figure \ref{f3-app} displays an example of evolution with variable number of links from $x4$ to $p2x2$.  

So far we have considered simple paths with no loops (also called off-pathways). The off-pathways can localize the dynamics in a small region of the configuration space wasting the evolution time. To escape from these traps, one may increase the path length $T$ in the hope of finding another path dominating the off-pathways, but probably another set of off-pathways would appear. 
Another strategy is to perturb the system, for example, by adding the transition rates $\mathcal{N}_{x\to p,s}$ to the original energy function $\mathcal{E}=-\sum_{t=1}^{T-1}N_p(t)$. However, we observe that in this case the off-pathways are very robust. The reason is that the off-pathways that maximize $\mathcal{N}_p$, maximize also the number of these transitions making the above perturbations ineffective. Table \ref{table-2} shows some examples of evolution in the presence of off-pathways.

\begin{table}
\begin{center}

\begin{tabular}{|c||c|c||c|c|c|}
  \hline

    $T$     & $\mathcal{N}_p$ &  $(g,\Delta)$ &
    $N_p^{off}$ &  $(g,\Delta)^{off}$ & off-pathway \\
  \hline

    $4$     & $31$  & $(1,1)$  & $31$ & $(1,1)$ & -  \\

  \hline

    $6$    & $58$  & $(3,1)$  & $58$ & $(3,1)$ & -  \\

  \hline

    $8$    & $86$  & $(24,1)$  & $86$ & $(24,1)$ & -  \\

  \hline

    $10$    & $113$  & $(48,1)$  & $114$ & $(24,1)$ & $\to loop$  \\

  \hline

    $12$    & $142$  & $(48,1)$  & $143$ & $(24,1)$  & $\to loop \to loop$ \\

  \hline

    $14$      & $169$  & $(372,1)$  & $172$ & $(120,1)$ & $\to loop \to loop \to loop$  \\

  \hline

\end{tabular}

\vskip 0.5cm

\caption{The total number of parallel two-links $\mathcal{N}_p$, degeneracy of the optimal paths $g$, and energy gap $\Delta$ obtained by an exhaustive search algorithm with local changes of type II. The optimal paths maximizing $\mathcal{N}_p$ connect tow  boundary configurations of $M=6$ links at shortest distance $D_{II}=4$. We compare the cases with and without off-pathways. An off-pathway which is a single loop appears for the first time at $T=10$. By increasing the path length $T$, we observe that more loops appear following each other.}\label{table-2}
\end{center}
\end{table}

\begin{figure}
\includegraphics[width=13cm]{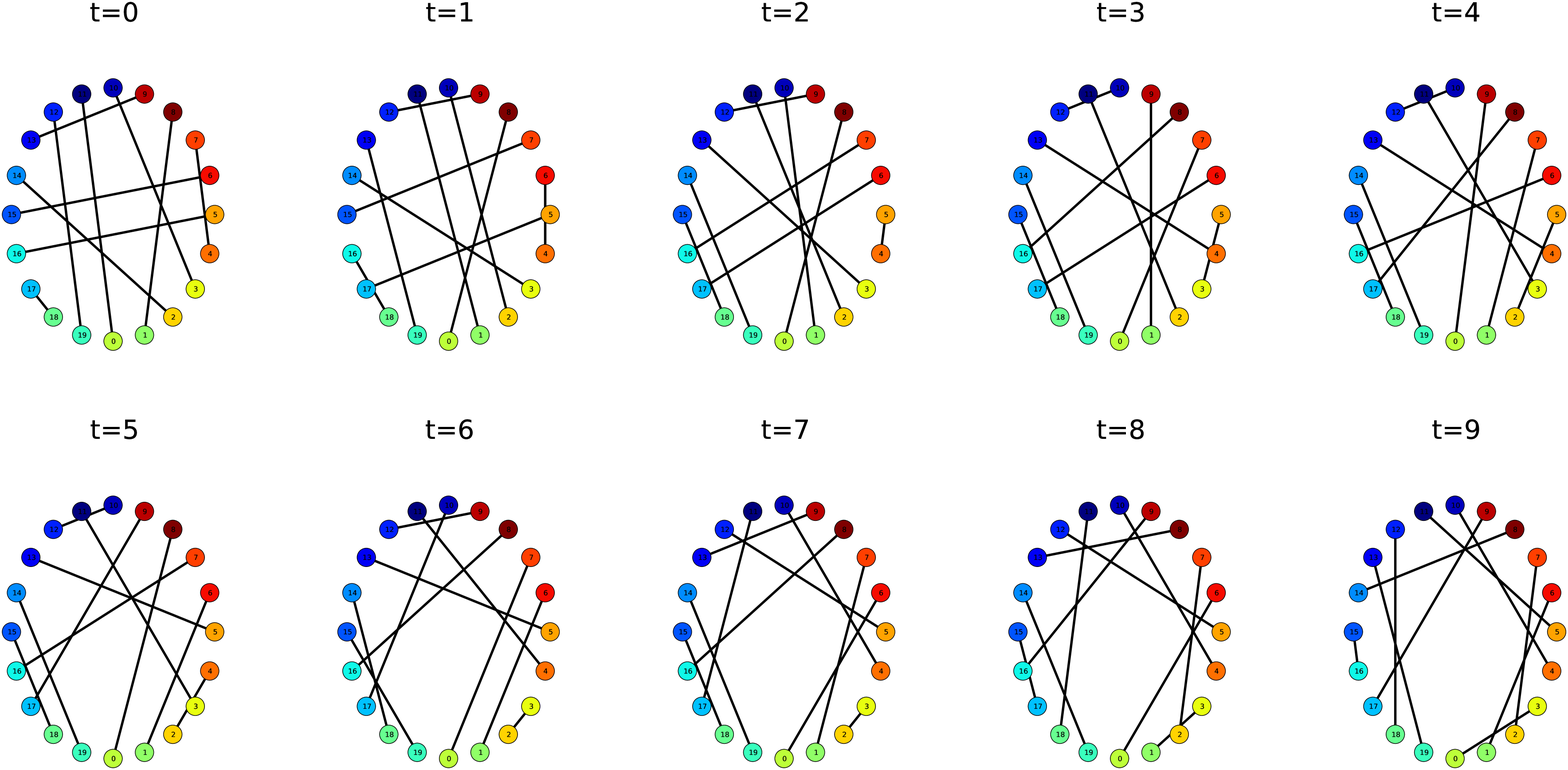} 
\includegraphics[width=13cm]{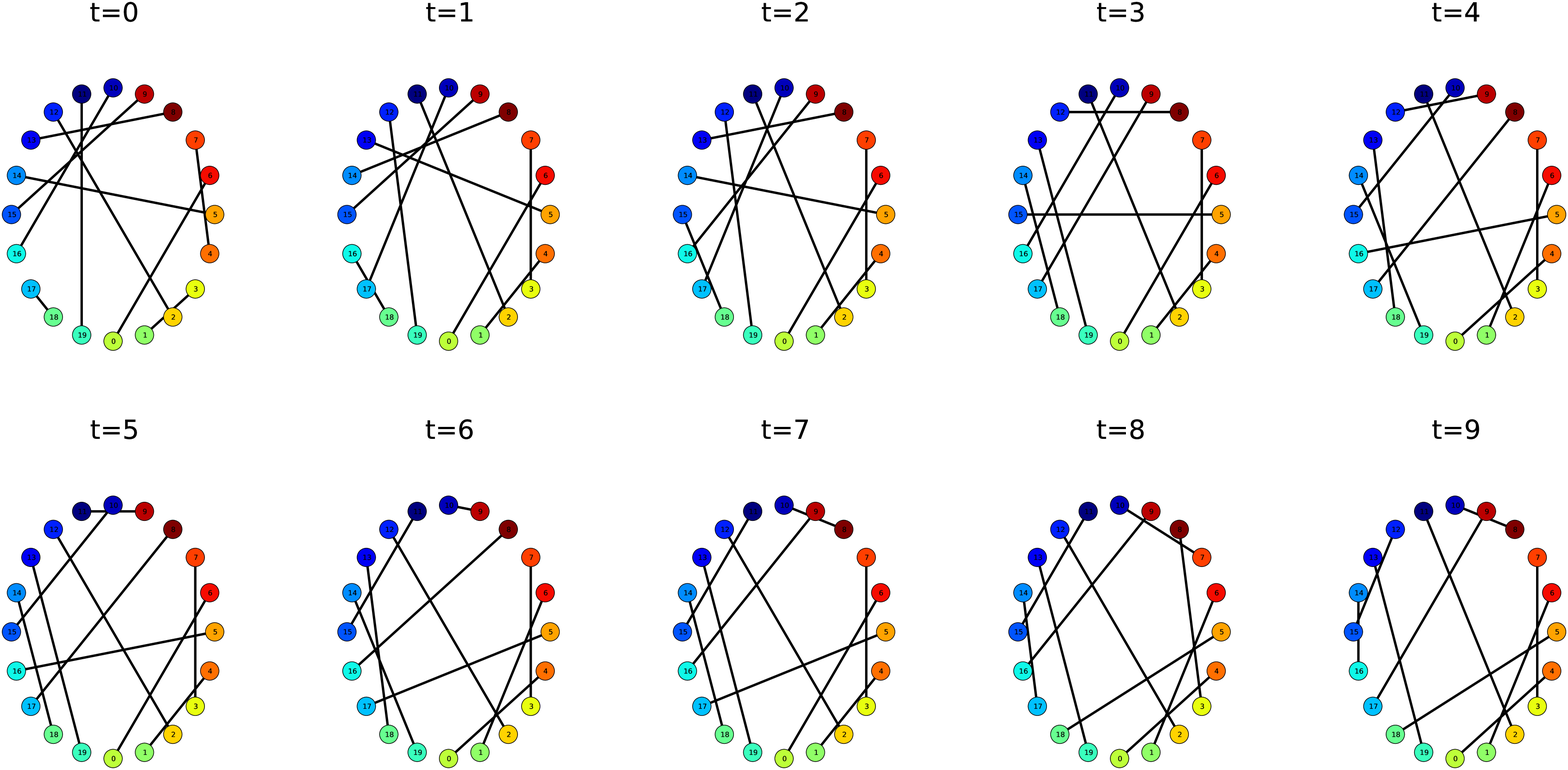} 
\includegraphics[width=13cm]{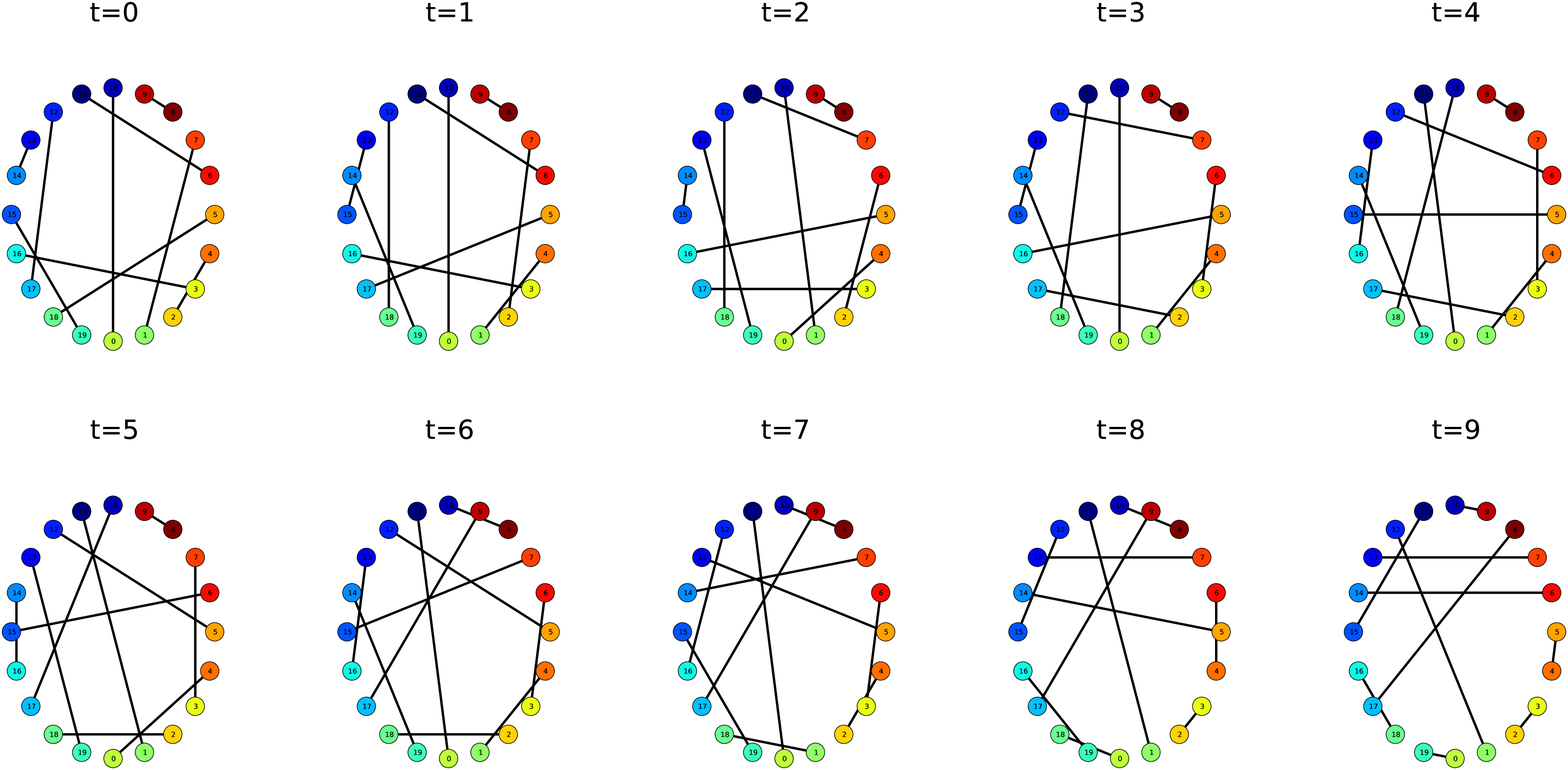} 
\caption{Evolution paths of $M=10$ links for $T=9$ coarse-grained steps from a random link configuration to another random configuration obtained by the approximate minimum evolution algorithm minimizing $\mathcal{E}=-\sum_{t=1}^{T-1}N_p(t)$. The endpoints on the chain start from $i=0$ (at the bottom of the circle) and increase to $i=2M-1$ in the counter-clockwise direction.}\label{f4-app}
\end{figure}

Figure \ref{f4-app} display the results we obtained by the approximate minimum-evolution algorithm for paths minimizing $\mathcal{E}=-\sum_{t=1}^{T-1}N_p(t)$. In this figure, we show instances of evolution paths between two random link configurations for a larger number of links $M=10$ and steps $T=9$.

\begin{thebibliography}{prsty}

\bibitem{MWT-structure-2014} A. Mashaghi, R.J. van Wijk and S.J. Tans, Structure {\bf 22}(9), 1227-1237 (2014).

\bibitem{MTM-pccp-2014} A. Mugler, S.J. Tans and A. Mashaghi,  Phys Chem Chem Phys. {\bf 16}(41):22537-44 (2014).

\bibitem{BC-pnas-1999} E. Bornberg-Bauer and H.S. Chan, Proc. Natl. Acad. Sci. {\bf 96}(19):10689-10694 (1999).
\bibitem{BKV-nature-2012} M.S. Breen, C. Kemena,	P.K. Vlasov, C. Notredame and F.A. Kondrashov, Nature {\bf 490} 535-538 (2012).

\bibitem{ZW-pnas-2015} B. Zhang and P.G. Wolynes, Proc. Natl. Acad. Sci. {\bf 112}(19):6062-6067 (2015).
\bibitem{JobDekker-nature-2012} E.P. Nora, B.R. Lajoie, E.G. Schulz, L. Giorgetti, I. Okamoto, N. Servant, T. Piolot,	 N.L. van Berkum, J. Meisig, J. Sedat, J. Gribnau, E. Barillot, N. Blüthgen,	 J. Dekker	 and E. Heard, Nature {\bf 485} 381-385 (2012).

\bibitem{CLW-pnas-2006} S.S. Cho, Y. Levy and P.G. Wolynes, Proc. Natl. Acad. Sci. {\bf 103}(3) 586-591 (2006).


\bibitem{FT-job-1963} H. Fisher and G.L. Thompson, in: Muth J.F, Thompson G.L. Industrial Scheduling, Prentice-Hall, Englewood (1963).

\bibitem{DRBZ-natphys-2011} S. Diehl, E. Rico, M.A. Baranov and P. Zoller, Nature Physics {\bf 7} 971-977 (2011).

\bibitem{R-pre-2013} A. Ramezanpour, Phys. Rev. E {\bf 87}, 060103(R) (2013). 


\bibitem{MP-epjb-2001} M. M\'ezard and G. Parisi,  Eur. Phys. J. B {\bf 20}, 217, 2001.

\bibitem{MM-book-2009} M. M\'ezard and A. Montanari, {\it Information, Physics, and Computation} (Oxford University Press, Oxford, 2009).

\bibitem{MZ-pre-2002}M. M\'ezard and R. Zecchina, Phys. Rev. E \textbf{66}, 056126 (2002).

\bibitem{MPZ-science-2002}M. M\'ezard, G. Parisi and R. Zecchina, Science \textbf{297}, 812 (2002).


\bibitem{steiner-prl-2008} M. Bayati, C. Borgs, A. Braunstein, J. Chayes, A. Ramezanpour, and R. Zecchina, Phys. Rev. Lett. {\bf 101}, 37208 (2008)

\bibitem{RRZ-epjb-2011} A. Ramezanpour, J. Realpe-Gomez, and R. Zecchina, Eur. Phys. J. B \textbf{81}, 327 (2011).

\bibitem{ABDZ-jstat-2013}F. Altarelli, A. Braunstein, L. Dall'Asta, and R. Zecchina, J. Stat. Mech. (2013) P09011.

\bibitem{YS-prl-2012} C. H. Yeung and D. Saad, Phys. Rev. Lett. {\bf 108}, 208701 (2012).

\bibitem{YSM-pnas-2013}C. H. Yeung, D. Saad, and K. Y. M. Wong, Proc. Natl. Acad. Sci. {\bf 110}(34), 13717 (2013).

\bibitem{ABRZ-prl-2011}F. Altarelli, A. Braunstein, A. Ramezanpour, and R. Zecchina, Phys. Rev. Lett. \textbf{106}, 190601 (2011).


\bibitem{LSS-prb-2008}C. Laumann, A. Scardicchio, and S. L. Sondhi,  Phys. Rev. B \textbf{78}, 134424 (2008).

\bibitem{AM-jstat-2011} E. Aurell and H. Mahmoudi, J. Stat. Mech. (2011) P04014.

\bibitem{LMOZ-pre-2014} A. Y. Lokhov, M. Mezard, H. Ohta, and L. Zdeborova,
Phys. Rev. E \textbf{90}, 012801 (2014).


\bibitem{KFL-inform-2001}F. R. Kschischang, B. J. Frey, and H. -A. Loeliger, IEEE Trans. Infor. Theory \textbf{47}, 498 (2001)

\bibitem{MR-RSC-2015}A. R. Mashaghi and A. Ramezanpour, RSC Adv. {\bf 5}(34), 51682 - 51689 (2015).

\bibitem{PKWT-nature-2007} F. J. Poelwijk, D. J. Kiviet, D. M. Weinreich, and S. J. Tans,
Nature {\bf 445}, 383 (2007).

\bibitem{BR-prl-2006}A. Braunstein and R. Zecchina, Phys. Rev. Lett. \textbf{96}, 30201 (2006).

\bibitem{QM-macro-2014}J. Qin and S.T. Milner, Macromolecules \textbf{47}(17), 6077 - 6085 (2014).


%% New

\bibitem{DSY-nature-2012} J.R. Dixon, S. Selvaraj, F. Yue, A. Kim, Y. Li, Y. Shen, M. Hu, J.S. Liu, and B. Ren, Nature 4{\bf 485}(7398) 376-380 (2012)

\bibitem{JCC-NAR-2012} D. Jost, P. Carrivain, G. Cavalli, and C. Vaillant, Nucl. Acids Res. {\bf 42} (15) 9553-9561 (2014)

\bibitem{PRDW-Nature-2014} B.D. Pope, T. Ryba, V. Dileep, F. Yue,  W. Wu, O. Denas,	D.L. Vera,	 Y. Wang,	 R.S. Hansen, T.K. Canfield, R.E. Thurman, Y. Cheng, G. Gülsoy, J.H. Dennis, M.P. Snyder, J.A. Stamatoyannopoulos, J. Taylor, R.C. Hardison, T. Kahveci, B. Ren, and D.M. Gilbert, Nature {\bf515} 402-405 (2014) 

\bibitem{DDK-science-2002} J. Dekker, K. Rippe, M. Dekker, and N. Kleckner, Science {\bf 295} (5558), 1306-1311 (2002)

\bibitem{BLW-JOVE-2010} N.L. Van Berkum, E. Lieberman-Aiden, L. Williams, M. Imakaev, A. Gnirke, L.A. Mirny, J. Dekker, and E. S. Lander, JoVE. {\bf 39} 1869 (2010)

\bibitem{DRA-GenRes-2006} J. Dostie, T.A. Richmond, R.A. Arnaout, R.R. Selzer, W.L. Lee, T.A. Honan, E.D. Rubio, A. Krumm, J. Lamb, C. Nusbaum, R.D. Green, and J. Dekker. Genome Res. {\bf 16}(10) 1299-309 (2006)

\bibitem{ZTS-naturegenet-2006} Z. Zhao, G. Tavoosidana, M. Sjölinder, A. Göndör, P. Mariano, S. Wang, C. Kanduri, M. Lezcano, K.S. Sandhu, U. Singh, V. Pant, V. Tiwari, S. Kurukuti, and R. Ohlsson, Nature Genetics {\bf 38} 1341- 1347 (2006)

\bibitem{SRB-PCB-2013} M. Di Stefano, A. Rosa, V. Belcastro, D. di Bernardo, C. Micheletti, PLoS Comput .Biol. {\bf 9}(3) e1003019 (2013)

\bibitem{PM-PRL-2013} O. Pulkkinen, and R. Metzler, Phys. Rev. Lett. {\bf 110}(19) 198101 (2013)

\bibitem{AH-PRL-2013}  A. Amitai and D. Holcman, Phys. Rev. Lett. {\bf 110} 248105 (2013)



\bibitem{F-pnas-2003}R. F. Fox, Proc. Natl. Acad. Sci. {\bf 100}(22), 12537?12538 (2003) 
\bibitem{CTW-JCP-2005} J. M. Carr, S. A. Trygubenko, and D. J. Wales, J. Chem. Phys. \textbf{122}, 234903 (2005).
\bibitem{AFR-JCP-2006} R. J. Allen, D. Frenkel, and P. R. ten Wolde, J. Chem. Phys. \textbf{124}, 024102 (2006).
\bibitem{MWZ-pre-2012} T. Mora, A. M. Walczak, and F. Zamponi, Phys. Rev. E \textbf{85}, 036710 (2012).
\bibitem{CC-RSC-2015}G. E. Crooks and D. Chandler, Phys. Rev. E, \textbf{64}, 026109 (2001).


\end{thebibliography}
\end{document}